\documentstyle[preprint,eqsecnum,aps]{revtex}
\newif\ifFigureInclude \FigureIncludefalse
\FigureIncludetrue

\catcode`@=11 
\def\lsim{\mathrel{\mathpalette\@versim<}}
\def\gsim{\mathrel{\mathpalette\@versim>}}
\def\@versim#1#2{\lower0.2ex\vbox{\baselineskip\z@skip
\lineskip\z@skip\lineskiplimit\z@\ialign{$\m@th#1
\hfil##\hfil$\crcr#2\crcr\sim\crcr}}}

\ifFigureInclude\input{epsf}
\else{\typeout{(Figure won't be included)}}\fi

\begin{document}
\draft
\preprint{$
\begin{array}{l}
\mbox{\bf hep-ph/9406252}\\[-3mm]
\mbox{\bf KEK--TH--376 }\\[-3mm]
\mbox{\bf KEK preprint 93--160 }\\[-3mm]
\mbox{\bf UT--659 }\\[-3mm]
\mbox{March~1994} \\[-3mm]
\mbox{\bf H}\\[1cm]
\end{array}
$
}

\title{Gluon and Charm Distributions in the Photon }
\author{K.~Hagiwara, M.~Tanaka and I.~Watanabe\cite{ISAMU}}
\address{Theory Group, KEK, Tsukuba, Ibaraki 305, Japan }
\author{T.~Izubuchi}
\address{Dept.\ of Physics, Univ.\ of Tokyo, Bunkyo-ku, Tokyo 113,
Japan }
\date{March 1994}
\maketitle
\begin{abstract}
  We study systematically the sensitivity of the photon structure
function
data on the gluon contents of the photon, by using the leading order
Altarelli--Parisi (AP) equations.
  Charm quark contribution is studied in the quark parton model and
by
using  the massive
quark AP equations of Gl\"uck, Hoffmann and Reya.
  The present photon structure function data are found to allow wide
range of
gluon distributions in the photon.
  We give a set of six scale-dependent parton distributions in the
photon
(WHIT1 to WHIT6), which have systematically different gluon contents.
  Sensitivity of the structure function at small $x$ and that of the
total
charm quark production cross section to the effective gluon
distribution are
discussed.
\end{abstract}
\section{Introduction}
  The deep structure of the photon has been expected to be calculable
perturbatively in QCD \cite{W}, except at small $x$ \cite{BBDORAG}.
  In practice, however, non-perturbative effects are found to be
significant
\cite{GR,DG} at present experiments, where the electromagnetic
structure of
Weizs\"acker--Williams quasi-real photon \cite{WW} has been measured
up to the
momentum transfer scale $Q^2$ = 100 GeV$^2$ in $e^+ e^-$ collision
experiments.
  Several parametrizations of the scale-dependent effective parton
distributions in the photon have been proposed; some are based on
plausible
dynamical assumptions \cite{GR,FKP,GS,ACFGP,GRV} and the others
\cite{DG,ACL}
have been obtained by fitting phenomenologically to the photon
structure
function data \cite{PLUTO1,PLUTO2,TASSO,JADE,TPC2G1,TPC2G2,AMY}.
  These parametrizations typically have similar quark distributions
which
are directly constrained by the structure function data.
  On the other hand, wildly different gluon distributions have been
proposed
since the electromagnetic structure of the photon is rather
insensitive to its
gluon content.

  Recently TRISTAN experiments \cite{AMYJET,TOPAZJET} have shown
evidence for
the effective gluon content of the photon in two-photon production of
high
transverse momentum ($p_T$) jets.
  The observed jet production cannot be understood without the
contribution
from gluons in the colliding photons, whereas it does not allow a
very hard
gluon distribution \cite{ACL} that is consistent with the structure
function
data.

  More recently, the TRISTAN experiments have reported evidence for
copious
production of charmed particles in the two-photon collision process,
in
their various decay modes and at various $p_T$ range
\cite{TOPAZC,VENUSC,AMYC}.
  The charm production rate has been found to be particularly
sensitive
to the gluon distribution in the photon \cite{DGod} and that the
present data
tend to prefer those parametrizations with large gluon content at
small $x$
($x\lsim 0.1$).

  The recent data on the jet and charm production in the two-photon
process
thus give us evidence and some quantitative information of the gluon
content
of the photon, but they are not yet accurate enough to determine
directly the
effective gluon distribution.
  It is therefore desirable to have a set of effective parton
distributions in
the photon with systematically different gluon distributions, so that
we can
learn more about the photon structure from these experiments.

  In this paper we study all the available photon structure function
data
\cite{PLUTO1,PLUTO2,TASSO,JADE,TPC2G1,TPC2G2,AMY,TOPAZ,VENUS,OPAL} at
4 GeV$^2
\leq Q^2 \leq$ 100 GeV$^2$ in the leading order of perturbative QCD
and, find
a new set of the effective scale-dependent parton distributions in
the photon,
named WHIT1 to WHIT6, which are all consistent with the present data
of
the photon
structure function and have systematically different gluon contents.
  We study carefully the charm quark contributions to the observed
structure
functions, which are evaluated by using the lowest order quark parton
model
matrix elements ($\gamma^* \gamma \rightarrow c {\bar c}$ and
$\gamma^* g \rightarrow c {\bar c}$) and the massive
Altarelli--Parisi (AP) equations \cite{GHR}.
  We find that the photon structure function has a sensitivity to the
gluon
distribution at small $x$, but that a careful analysis is needed to
determine
experimentally the photon structure in this region.
  Predictions are also given for the total charm quark pair
production cross
section in the two-photon collision process at $e^+ e^-$ colliders.

  We note here that the next-to-leading order correction to the
massless
inhomogeneous AP equations has been known for a while
\cite{GR,BB,GRV0,FP}, and some phenomenological analyses
\cite{FKP,GS,ACFGP,GRV} were performed at this level.
  Recently, the complete next-to-leading order correction has been
obtained
for the massive quark production process \cite{LRSV}.
  We work in the leading order of QCD, nevertheless, since errors in
the
experimental data as well as the theoretical uncertainties associated
with the
gluon contents of the photon are so large that the leading order
approach is
more suited to discuss them systematically.

  The paper is organized as follows.
  In Sec.~\ref{MODEL}, we review the AP equations that govern the
effective parton
distributions in the photon and introduce the notion of `valence' and
`sea'
components of the quark distributions.
  We discuss our parametrizations of the initial quark and gluon
distributions
and study charm contributions to the structure function.
  In Sec.~\ref{FITTOTHEDATA}, we introduce all available photon
structure
function data, give a criterion to remove several low $x$
experimental data,
and then make a global fit of the initial light-quark distribution
functions
by using the leading order inhomogeneous AP equations.
  The fit is repeated by systematically changing the magnitude and
the shape
of the initial gluon distribution.
  The minimal $\chi^2$ of the fit as well as the distribution of the
deviation
of each
data point from the best fit curve is examined carefully.
  Six effective parton distributions which have systematically
different
initial gluon distributions, WHIT1 to WHIT6, are then introduced
and examined.
  In Sec.~\ref{EFFECTIVE}, we discuss effective heavy quark
distributions
in the photon by using the quark parton model (QPM) and the massive
AP equations.
  In Sec.~\ref{CHARMPROD},
predictions are  given for the total charmed particle production
cross
section in the two-photon process at $e^+e^-$ colliders.
Sec.~\ref{CONCLUSIONS} summarizes our findings.
  The details on the numerical methods that we use in order to solve
the AP
equation with and
without charm quark mass effects are give in Appendix~\ref{APXA},
and the parametrizations of our effective parton
distributions in the photon are described in
Appendix~\ref{Parametrization}.
\section{Model}
\label{MODEL}
In this section we explain the framework adopted in this work to
calculate
the effective parton distribution functions in the photon and the
photon
structure function $F_2^\gamma(x,Q^2)$.

\subsection{Inhomogeneous Altarelli--Parisi equations}

  In the $Q^2$ and $x$ region where the lightest $n_f$-flavor quarks
can be produced,
we have $n_f$ quark distributions and $n_f$ anti-quark distributions
in addition
to the gluon distribution in the photon.
  These are denoted by $q_i(x,Q^2)$, $\bar q_i(x,Q^2)$ ($i$ = 1 to
$n_f$), and
$g(x,Q^2)$ respectively.
  Apparently, the relation $q_i(x,Q^2) = \bar q_i(x,Q^2)$ holds for
each
flavor.

  The $Q^2$ evolution of these parton distributions is described by
the
inhomogeneous Altarelli--Parisi (AP) equations  in the leading
logarithmic
approximation \cite{DJSWW}.
  For massless $n_f$-flavor case the AP equations can be written as
follows:
\begin{mathletters}
\label{FIAPE}
\begin{eqnarray}
\frac{dq_i(x,Q^2)}{dt}&=&
      \frac{\alpha}{2\pi}e_i^2 P_{q\gamma}(x)+
      \frac{\alpha_s(Q^2)}{2\pi}\left[P_{qq}(x)\otimes q_i(x,Q^2)+
                                      P_{qg}(x)\otimes
g(x,Q^2)\right]\,,
\label{FIAPEQ}\\
\frac{dg(x,Q^2)}{dt}&=&
      \frac{\alpha_s(Q^2)}{2\pi}\left[ 2 P_{gq}(x)\otimes
                                      \sum_{i=1}^{n_f} q_i(x,Q^2)+
                                      P_{gg}(x,n_f)\otimes
g(x,Q^2)\right]\,,
\label{FIAPEG}
\end{eqnarray}
\end{mathletters}
where $i$ = 1 to $n_f$, $t = \log Q^2/\Lambda^2$ with $\Lambda$ being
the QCD scale, and $P_{ij}\/$'s are the
parton splitting functions \cite{AP} whose explicit forms are given
in
Eq.~(\ref{eq:mlSPLITF}) of Appendix~\ref{APXA}.
  The convolution integral is defined as $a(x)\otimes b(x) =
\int_x^1 dy/y\, a(x/y) b(y)$.

  As we show in the latter sections, the charm quark contribution to
the
photon structure function can be most conveniently calculated from
the lowest
order quark parton model processes ($\gamma^* \gamma \rightarrow c
\bar{c}$
and $\gamma^* g \rightarrow c \bar{c}$) at present energies ($Q^2
\lsim$ 100
GeV$^2$), by excluding the charm quark distribution in the photon.
  Only at higher $Q^2$ we introduce the effective charm quark
distribution
that evolves according to the massive $n_f$ = 4 AP equations of
Ref.\cite{GHR}.
At very high $Q^2$, we may neglect the charm quark mass and employ
the
massless inhomogeneous AP equations of Eq.~(\ref{FIAPE}) with
$n_f$=4.
  The matching of the quark parton model description with the
effective heavy quark distribution in the massive AP equations
is discussed in Sec. ~\ref{EFFECTIVE}.

  Hence in the analysis of the present structure function data that
probe the photon structure up to $Q^2\sim 100 {\rm GeV}^2$, we
introduce
only three light-quark distributions ($n_f$ = 3).
  Furthermore, in order to find a plausible parametrization of these
quark
distributions for the fit, we find it convenient to introduce the
notion of
`valence' and `sea' quark distributions.
  The `valence' quarks are produced by the photon and the `sea'
quarks originate from the gluons.
  According to these definitions, the valence and sea quark
distributions are
written in terms of the original quark distributions $q_i$'s:
\begin{mathletters}
\begin{eqnarray}
q_v(x,Q^2)&=&2\sum_{i=1}^3\frac{e_i^2\langle e^2\rangle-\langle
e^2\rangle^2}
                               {\langle e^4\rangle-\langle
e^2\rangle^2}
                          q_i(x,Q^2)\,,
\label{VAL}\\
q_{sea}(x,Q^2)&=&2\sum_{i=1}^3\frac{\langle e^4\rangle-e_i^2\langle
e^2\rangle}
                                   {\langle e^4\rangle-\langle
e^2\rangle^2}
                              q_i(x,Q^2)\,,
\label{SEA}
\end{eqnarray}
\end{mathletters}
where $\langle e^2\rangle$ = 2/9 and $\langle e^4\rangle$ = 2/27 for
$n_f$ = 3.
  Note that the singlet and non-singlet quark distributions,
$q_{S}(x,Q^2)$ and
$q_{NS}(x,Q^2)$, respectively, are
related to our valence and sea quark distributions by
\begin{mathletters}
\begin{eqnarray}
q_S(x,Q^2)    & \equiv & 2 \sum_{i=1}^3 q_i(x,Q^2)\,, \nonumber \\
              & =      & q_v(x,Q^2)+q_{sea}(x,Q^2)\,,
\label{SINGLET} \\
q_{NS}(x,Q^2) & \equiv & 2 \sum_{i=1}^3
                         \left[ e_i^2 - \langle e^2 \rangle \right]
                         q_i(x,Q^2)\,, \nonumber \\
              & =      & \left[\frac{\langle e^4\rangle}{\langle
e^2\rangle}
                               - \langle e^2 \rangle\right]
                         q_v(x,Q^2)\,.
\end{eqnarray}
\end{mathletters}

  The photon structure function $F_2^\gamma(x,Q^2)$ can be written in
terms of
$q_v(x,Q^2)$ and $q_{sea}(x,Q^2)$ as
\begin{eqnarray}
F_2^\gamma(x,Q^2) & \equiv & 2 x \sum_{i=1}^{n_f} e_i^2 q_i(x,Q^2)\,,
                                                      \nonumber \\
   & = & x\left[\frac{\langle e^4\rangle}{\langle
e^2\rangle}q_v(x,Q^2)+
                         \langle e^2 \rangle q_{sea}(x,Q^2)\right]
         + \mbox{heavy quarks}\,.
\end{eqnarray}
  Heavy quark contributions will be discussed in
subsection~\ref{CHARM}.
  When we neglect small mass differences among the light three
flavors, we can
express the $u$, $d$ and $s$ distributions in terms of $q_v$ and
$q_{sea}$:
\begin{mathletters}
\label{UANDD}
\begin{eqnarray}
 u(x,Q^2)         &=& \frac{1}{3} q_v(x,Q^2)+\frac{1}{6}
q_{sea}(x,Q^2)\,,\\
 d(x,Q^2)=s(x,Q^2)&=& \frac{1}{12}q_v(x,Q^2)+\frac{1}{6}
q_{sea}(x,Q^2)\,.
\end{eqnarray}
\end{mathletters}

  The AP equations of Eq.~(\ref{FIAPE}) with $n_f=3$ can be rewritten
in terms of the valence-quark,
 the sea-quark and the gluon distributions:
\begin{mathletters}
\label{IAPE}
\begin{eqnarray}
\frac{dq_v(x,Q^2)}{dt}&=&
                       \frac{\alpha}{2\pi}\langle e^2\rangle
P_{q\gamma}(x)+
                       \frac{\alpha_s(Q^2)}{2\pi}P_{qq}(x)\otimes
q_v(x,Q^2)\,,
\label{IAPEV}\\
\frac{dq_{sea}(x,Q^2)}{dt}&=&\frac{\alpha_s(Q^2)}{2\pi}
                       \left[ 2\cdot 3 P_{qg}(x)\otimes g(x,Q^2)+
                             P_{qq}(x)\otimes
q_{sea}(x,Q^2)\right]\,,
\label{IAPESEA}\\
\frac{dg(x,Q^2)}{dt}&=&\frac{\alpha_s(Q^2)}{2\pi}
\left[P_{gq}(x)\!\otimes\!\left\{\!q_v(x,Q^2)\!+\!
q_{sea}(x,Q^2)\!\right\}\!+\!
                             P_{gg}(x,3)\!\otimes\!
g(x,Q^2)\right]\!.
\label{IAPEG}
\end{eqnarray}
\end{mathletters}
 It is now clearly seen that the valence-quarks are produced by the
photon,
the sea-quarks are produced by the gluon, while the gluon is produced
by
the valence-quarks, the sea-quarks and the gluon itself.
  Once a set of initial parton distributions at $Q^2=Q_0^2$ is given,
we can
predict the parton distributions at any $Q^2(>Q_0^2)$ by solving the
above
equations numerically.
  The numerical methods which we use to solve these equations are
 explained in Appendix~\ref{APXA}.

\subsection{Initial parton distributions}
\label{IPD}

  To solve the AP equations of Eq.~(\ref{IAPE}), we have to specify a
set of initial
parton distributions at $Q^2 = Q_0^2$.
  All the non-perturbative features of the photon structure are
included in
these initial conditions.
  We use $Q_0^2$ = 4 GeV$^2$ throughout our analysis in order that
our
perturbation approximation works well.

  As an initial valence-quark distribution, we take the following
functional
form:
\begin{equation}
xq_v(x,Q_0^2)/\alpha=A_v x^{B_v} (1-x)^{C_v}/B(B_v+1,C_v+1)\,,
\label{INPUTV}
\end{equation}
where $A_v$, $B_v$ and $C_v$ are the free parameters which will be
fitted to
the experimental data, and $B(\alpha,\beta)$ is the beta function
that ensures
the normalization,
\begin{equation}
\langle xq_v(x,Q_0^2)\rangle/\alpha \equiv \int_0^1
dx\,xq_v(x,Q_0^2)/\alpha=A_v\,,
\label{NORV}
\end{equation}
for the energy fraction $\langle xq_v(x,Q_0^2)\rangle$ of the
valence-quarks in
the photon.

  As for the initial gluon distribution, we adopt the simple form
\begin{equation}
xg(x,Q_0^2)/\alpha=A_g (C_g+1) (1-x)^{C_g}\,,
\label{INPUTG}
\end{equation}
with two parameters, $A_g$ and $C_g$.
  Again the normalization factor is chosen such that
\begin{equation}
  \langle xg(x,Q_0^2)\rangle/\alpha=\int_0^1\,dx xg(x,Q_0^2)/\alpha
=A_g
\end{equation}
  The present structure function data are not accurate enough to
determine the
gluon parameters $A_g$ and $C_g$.
  We therefore perform fit by the valence-quark parameters, $A_v$,
$B_v$
and $C_v$, by varying systematically the normalization ($A_g$) and
the shape
($C_g$) of the initial gluon distribution.

  Before starting the fit to the data, we discuss plausible range of
the gluon
distribution parameters that we should explore.
  We obtain  constraints on the ratio of the gluon energy
fraction ($A_g$) to the valence-quark energy fraction ($A_v$) as
follows
\cite{GS}.
  At sufficiently low momentum transfer scale ($Q^2 \lsim$ 0.5
GeV$^2$), only
the long wavelength components of the photon are probed and the
quark-antiquark pair produced from the photon undergoes
non-perturbative soft
QCD dynamics that resembles the one which makes the quark-antiquark
pair
form the low-lying vector boson.
  The photon structure is then expected to have components similar to
those of
the vector bosons, in particular the $\rho$ meson that couples
strongly to the
photon.
  Although we do not know the structure of $\rho$, we expect the soft
QCD
dynamics to be insensitive to the total spin of the system and that
it may be similar to the observed $\pi$ structure \cite{ABFKW}.
  If the photon had only this soft component, its deep structure
should also
be similar and we expect
\begin{equation}
\frac{\langle xg(x,Q_0^2)\rangle_{\mbox{\scriptsize `$\rho$'}}}%
     {\langle xq_v(x,Q_0^2)\rangle_{\mbox{\scriptsize `$\rho$'}}}
\sim
\frac{\langle xg(x,Q^2_0)\rangle_{\pi}}{\langle
xq_v(x,Q_0^2)\rangle_{\pi}}
\sim 1\,.
\label{CONSTV}
\end{equation}
  In fact this ratio is common in the nucleon structure
as well \cite{PHDLNCL,GHR}
and we can regard this ratio as an universal one from soft QCD
dynamics.

  The photon, however, differs from the vector boson in that it is a
source of
a quark-pair with an arbitrary short wave-length.
  As the momentum transfer scale grows ($Q^2 \gsim$ 0.5 GeV$^2$), one
is more
and more sensitive to these short wave-length components which
dominate the
photon structure at asymptotically high $Q^2$.
  Although the transition from the regime where the vector meson-like
component dominates to the regime where the short wave-length
component
dominates is gradual and it is governed by the non-perturbative
dynamics, we
may infer the effect of the latter component from its asymptotic
behavior that
can be calculated perturbatively.
  In particular, for the ratio of the gluon to the valence-quark
energy fraction,
we expect
\begin{equation}
\frac{\langle xg(x,Q^2)\rangle}{\langle xq_v(x,Q^2)\rangle}
\Bigg|_{Q^2\rightarrow\infty}
=\frac{3616}{10611}\simeq\frac{1}{3}\,,
\label{CONSTA}
\end{equation}
for three light quark flavors ($n_f$ = 3).
  At the momentum transfer scale $Q_0^2$ = 4 GeV$^2$, it is hence
natural to
expect the ratio to lie somewhere between the two extremes
Eq.(\ref{CONSTV}) and
(\ref{CONSTA}):
\begin{equation}
\frac{1}{3}\lsim
\frac{\langle xg(x,Q_0^2)\rangle}{\langle xq_v(x,Q_0^2)\rangle}
=\frac{A_g}{A_v}
\lsim 1\,.
\label{CONST}
\end{equation}
  We shall see in the next section that the valence-quark fraction
$A_v$ is
determined to be about unity by the experimental data of
$F_2^\gamma(x,Q^2)$.
  We will hence examine the parameter range  $1/3 \lsim A_g \lsim 1$
for the
gluon energy fraction.

  Finally, we note that the sea-quark distribution is intimately
related to
the gluon distribution and that one cannot choose them independently.
  Although the sea-quark distribution is in principle observable from
the small
$x$ behavior of the photon structure function, we find that the
present
experimental
determination of the small $x$ structure of the photon suffers from
an uncertainty
associated with the unfolding technique  adopted by most
experiments \cite{UNFOLD}: this will be discussed in
Sec.~\ref{FITTOTHEDATA}.
  We therefore estimate the input sea-quark distribution by using the
quark
parton model cross section for the process $\gamma^* g \rightarrow
q \bar{q}$:

\begin{equation}
xq_{sea}(x,Q_0^2)= 3 \,\frac{\alpha_s(Q_0^2)}{2\pi} \int_{ax}^1 dy\,
                   w\left( \frac{x}{y},\frac{m^2}{Q_0^2}\right)
\,g(y,Q_0^2)\,,
\label{INPUTS}
\end{equation}
where $a=1+4m^2/Q_0^2$ and
\begin{eqnarray}
w(z,r) =
  &z&\Biggl[\beta\left\{-1+8z(1-z)-4rz(1-z)\right\}
\nonumber \\
              & &\quad+\left\{z^2+(1-z)^2+4rz(1-3z)-8r^2z^2\right\}
                       \log \frac{1+\beta}{1-\beta}\Biggr]
\label{BH}
\end{eqnarray}
with $\beta = \sqrt{1-4rz/(1-z)}$.
  The sea-quark mass $m$, which is taken to be common for the three
light
flavors, $m_u$ = $m_d$ = $m_s$ = $m$, plays the role of the cut-off
and we
choose it to be 0.5 GeV.
  Here and throughout our analysis we adopt the leading order form of
the QCD
running coupling constant
\begin{equation}
 \frac{\pi}{\alpha_s(Q^2)} = \frac{25}{12} \ln\frac{Q^2}{\Lambda_4^2}
                           - \frac{1}{6}   \ln\frac{Q^2}{4 m_b^2}
                             \Theta(Q^2-4 m_b^2) \,,
\label{QCDRUN}
\end{equation}
with $\Lambda_4$ = 0.4 GeV and $m_b=5$ GeV.
  Note that the effective number of quark flavors that governs the
running of
the coupling constant is chosen independently of the number $n_f$ of
massless
quark flavors in the AP equations of Eq.~(\ref{FIAPE}).
  An accurate prescription for the choice of the effective number of
flavors
is found only in the next-to-leading order level
\cite{LRSV}.

  We remark here that the above prescription leads naturally to a
larger
sea-quark input as the gluon input is enhanced.
  In particular, we find for the energy fraction ratio that
\begin{equation}
\frac{\langle xq_{sea}(x,Q_0^2)\rangle}{\langle xg(x,Q_0^2)\rangle}
\sim 0.12
\end{equation}
holds almost independently of the input gluon parameters $A_g$ and
$C_g$
in the region which we will discuss.
  The ratio increases with decreasing light-quark mass, and reaches
0.2 at
$m \sim$ 0.3 GeV.
  The ratio is about 0.3 in the parametrization of the $\pi$
structure
\cite{ABFKW}, and its perturbative asymptotic value is calculated to
be 0.16 for $n_f$ = 3.
  Our sea-quark input is hence rather conservative for a given input
gluon
distribution.

\subsection{Charm contribution}
\label{CHARM}

  The charm quark cannot be incorporated into the massless AP
equations
in the region
of moderate $Q^2$, say $Q^2 \le $ 100 GeV$^2$, which has so far been
probed by
experiments.
  We should take into account the quark mass effect by using the
massive-quark
AP equation of  M.~Gl\"uck, E.~Hoffmann, and E.~Reya \cite{GHR}, and
more accurately by
incorporating the full next-to-leading order corrections \cite{LRSV}.
  We find by comparing with the results of the leading order
massive-quark
AP equations that the charm quark contribution to the photon
structure function is well approximated by the sum of the
contributions from
the quark parton model processes $\gamma^* \gamma \rightarrow c
\bar{c}$ and
$\gamma^* g \rightarrow c \bar{c}$  at $Q^2 \le $ 100 GeV$^2$.
Beyond $Q^2\sim 100{\rm GeV^2}$ the radiation of gluons off charm
quarks is
no longer negligible and, we should solve the massive-quark
inhomogeneous
AP equations.
At large enough $Q^2$, the charm quark mass effect to the
$Q^2$-evolution
would become
negligible and we can use the massless AP equations of
Eq.~(\ref{FIAPE})
with $n_f$ = 4.
The matching of the distributions should then be made at
appropriately large $Q^2$.
We will find that the charm-quark mass effects are not negligible
even at
$Q^2\sim 100 {\rm GeV^2}$.
With the same criterion, the bottom quark contribution can be
estimated by the lowest order process $\gamma^* \gamma \rightarrow b
\bar{b}$
and $\gamma^* g \rightarrow b \bar{b}$ up to about $Q^2 \sim$ 1000
GeV$^2$,
above which we may introduce the effective b-quark distribution that
follows
the massive $n_f$ = 5 AP equations.
  More accurate quantitative treatment \cite{LRSV} will become useful
in the
future when both the quark and gluon distributions are measured
accurately
from experiments.

The charm-quark contributions to the photon structure function are
thus
calculated by
the quark parton model at $Q^2 < 100 {\rm GeV^2}$.
The contribution of the direct process ($\gamma^* \gamma \rightarrow
c \bar{c}$) is given by
\begin{equation}
F_{2,c}^\gamma(x,Q^2)|_{\rm direct}=3 \frac{\alpha}{\pi} e_c^4\,
                                      w\!\left(
x,\frac{m_c^2}{Q^2}\right)\,,
\label{CD}
\end{equation}
where $e_c = 2/3$ is the charm-quark electric charge and the function
$w(x,r)$ is
given in Eq.~(\ref{BH}).
  In our numerical analysis, we take $m_c$ = 1.5 GeV.
  For the resolved process ($\gamma^* g \rightarrow c \bar{c}$) , we
have
\begin{equation}
F_{2,c}^\gamma(x,Q^2)|_{\rm resolved}=\frac{\alpha_s(Q^2)}{2\pi}
e_c^2\,
     \int_{ax}^1 dy\, w\left(
\frac{x}{y},\frac{m_c^2}{Q^2}\right)\,g(y,Q^2)\,,
\label{CR}
\end{equation}
where $a = 1+ 4 m_c^2 / Q^2$, and the gluon distribution $g(x,Q^2)$
is given
by solving the massless $n_f$ = 3 AP equations of Eq.~(\ref{IAPE})
with
the initial parton distributions of Eqs.~(\ref{INPUTV}),
(\ref{INPUTG}) and
(\ref{INPUTS}).

  The validity of our simple quark parton model calculation depends
on how
much the
gluon emission by the charm quark distorts the effective charm quark
distribution in the photon.
  The magnitude of this effect can be studied by using the massive AP
equations
for the charm quark and is presented in Sec.~\ref{EFFECTIVE}.

\section{Fit to the data}
\label{FITTOTHEDATA}
\subsection{Data}

  In order to find good initial parton distributions at the energy
scale
$Q_0^2$ = 4 GeV$^2$, we refer to all the available experimental data
of the
photon structure function at $Q^2 > Q_0^2$.
  In our analysis we use the data obtained by 8 groups
at PETRA, PEP, TRISTAN and LEP $e^+ e^-$ colliders  which are listed
in
Table~\ref{TB1}.

  We note here that not all the experimental data points are taken
into
account in our fit.
  First, we do not use the data at $\langle Q^2 \rangle$
lower than 4.0 GeV$^2$.
  Second, we accept only those data points at small $x$ where the
following inequality holds:
\begin{equation}
     x_{\mbox{\footnotesize lower edge of the bin}} \quad
     >  \quad \frac{\langle Q^2 \rangle}{\langle Q^2 \rangle +
       (W_{\rm vis}^{\rm max})^2}\,.
\label{3.1} \end{equation}
  Here $W_{\rm vis}^{\rm max}$ is the experimental cut on the visible
invariant mass of the final hadron system.
  Since those data points that violate the condition Eq.(\ref{3.1})
are obtained at near the boundary of the
experimental acceptance and since the sea-quark contribution to the
structure
function can be rather
singular at the low $x$ region, they may suffer from large systematic
uncertainties in the unfolding procedure \cite{UNFOLD}.
  As a result of the above two requirements, 47 data points are
retained in
our fitting, which are all listed in Table~\ref{TB1}.

\subsection{Fit}
\label{FIT}
  By fitting our theoretical predictions for the photon structure
function to
these experimental data, we tune the parameters of the initial
valence-quark
distribution $A_v$, $B_v$ and $C_v$.
  We repeat the fit by varying the initial gluon distribution
parameters $A_g$
and $C_g$ systematically,
while keeping the strong coupling constant and the charm quark mass
fixed
at $\Lambda_4=0.4{\rm GeV}$ and $m_c=1.5{\rm GeV}$,
respectively.
  In particular, we examine the case with $A_g$ = 0.5, 1 and 1.5
systematically
by changing $C_g$ and find
little sensitivity of the structure function data to the shape
parameter
$C_g$.
  The fit results for arbitrarily chosen 12 cases are summarized in
Table~\ref{TB2}. Although there is a tendency in the data that
prefers small
$A_g$ (small gluonic energy fraction) and small $C_g$ (hard gluon
distribution),
it is caused by a few data points at small $x$ with relatively
large deviations from the best fit curve, as we will discuss below.
We choose three representative $C_g$ values ($C_g$=3, 9, 15) for each
of the normalizations $A_g$=0.5 and 1 that are consistent with the
ansatz
Eq.~(\ref{CONST}).
These gluon inputs are named as WHIT1 to WHIT6, respectively, as
shown in
Table~\ref{TB3}.

  Fig~\ref{F3.1} illustrates the matching of the data and the
theoretical curves,
and Fig.~\ref{F3.2} shows the distribution of the deviation of each
data point
from the best fit value, $(F_2(x)_{\rm fit}-F_2(x)_{\rm data})
/ \sigma(F_2(x)_{\rm data})$.
  As can be seen from Fig.~\ref{F3.1} and Fig.~\ref{F3.2},
all of WHIT1 to WHIT6 gluon distributions give similar quality of
fits to the
photon structure function data.
The mild dependence of the best fit $\chi^2$ value
on the choice of the initial gluon distribution parameters as shown
in
Table~\ref{TB2} is a consequence of a few data points at
the lowest $x$ bin that satisfy the criterion Eq.~(\ref{3.1}):
see Fig.~\ref{F3.2}.
In Fig.~\ref{F3.2}, deviations of those data points that are removed
from the
fit by the criterion Eq.~(\ref{3.1}) are also indicated by dashed
lines.
The large $A_g$ large $C_g$ gluon distributions lead to a significant
rise
in the structure function at small $x$, and a naive integration of
the
structure function in the given $x$-bin tends to be very sensitive to
the lower edge of the lowest $x$-bin as is seen in Fig.~\ref{F3.1}.
After imposing the selection criterion Eq.~(\ref{3.1}), such
sensitivity to the
very low $x$ behavior of the structure function is almost completely
lost,
as can be seen from the deviations of the data points that are
connected by the solid lines in Fig.~\ref{F3.2}.
In view of the relatively large theoretical uncertainty in simulating
hadronic events at small $x$, we conclude
that the present experimental data on the photon structure function
have poor sensitivity to the gluonic content of the photon.
  The normalization of the valence-quark distribution $A_v$ is found
to be
roughly 1, regardless of the difference in the sea-quark contribution
that
depends strongly on our gluon inputs.

  We find from Table~\ref{TB2} that the best fit values of the
initial valence-quark
parameters are
almost the same for different $C_g$'s for a common $A_g$.
  Hence we introduce a `standard' set of the valence-quark parameters
for each
$A_g$; {\em i.e.}
\begin{mathletters}
\label{STANDARD}
\begin{eqnarray}
 A_v = 0.94 \ , \quad B_v = 0.50 \ , \quad C_v = 0.25 \
& &\mbox{\ \ \ for $A_g = 0.5$}\,,
 \label{STANDARD1} \\
  A_v = 0.89 \ , \quad B_v = 0.70 \ , \quad C_v = 0.45 \
& &\mbox{\ \ \ for $A_g = 1.0$}\,.
 \label{STANDARD2}
\end{eqnarray}
\end{mathletters}
We calculate the $\chi^2$ values for various $C_g$'s by fixing the
normalization
$A_g$ and
the associated valence-quark inputs as above, and the result is
summarized in
Table~\ref{TB3}.
  The $C_g$ dependence of the resulting $\chi^2$ is also presented in
Fig.~\ref{F3.3}.
The minimal $\chi^2$ values as obtained in Table~\ref{TB2} by tuning
the valence
quark parameters are also shown by large symbols.
The tuning of the valence-quark parameters do not improve the fit
much.
We therefore adopt the common valence-quark input of
Eq.~(\ref{STANDARD1})
for the sets WHIT1 to
WHIT3, while that of Eq.(\ref{STANDARD2}) for the sets WHIT4 to
WHIT6,
and hereafter we call the parton distributions with these
inputs as WHIT1 to WHIT6.
  The slightly small valence-quark contributions of
Eq.(\ref{STANDARD2}) at small
$x$ partially compensates for the larger sea-quark contributions
associated with
the large gluon inputs of WHIT4 to WHIT6.

\subsection{Gluon distribution}
\label{DISCUSSIONS}

  As we described above, we present a set of six effective parton
distributions
in the photon with systematically different gluon contents.
We show in Fig.~\ref{F4.1} all the gluon distributions (WHIT1 to
WHIT6) at
three momentum transfer scales, $Q^2$=4, 20 and 100 GeV$^2$.
The area under the curves at $Q^2$= 4GeV$^2$ is given by the
normalization
$A_g=0.5$ for WHIT1 to WHIT3, and $A_g=1.0$ for WHIT4 to WHIT6.
The shape of the distribution becomes softer as we move from WHIT1 to
3,
and from WHIT4 to 6 in each set.
The huge difference in the initial gluon
distributions tends to diminish at higher $Q^2$, as expected from the
asymptotic behavior of the solution of the inhomogeneous AP
equations.

  Also shown in Fig.~\ref{F4.1} for comparison are the effective
gluon distributions
 of GRV\cite{GRV}, DG \cite{DG} and LAC1 \cite{ACL}
at the three momentum transfer scales.
We note that in the $x$ and $Q^2$ range as shown in Fig.~\ref{F4.1},
our WHIT1
effective gluon distribution is similar to the gluon distribution of
GRV
\cite{GRV},
while WHIT6 gluon distribution behaves similarly to that of LAC1
\cite{ACL}.
It should be noted, however, that LAC1 parametrization for the
effective gluon distribution is more singular at $x \rightarrow 0$
than that of
WHIT6, which results in a very large
energy fraction $\langle x g(x,Q_0^2) \rangle / \alpha$ = 2.37 at
$Q_0^2$ = 4
 GeV$^2$, in conflict with our ansatz Eq.~(\ref{CONST}) with $A_v
\sim 1$.
Accurate measurements of the photon structure functions at very small
$x$
as well as the high energy behavior of the charm and mini-jet
production
cross section at $\gamma\gamma$ and $\gamma p$ collisions will be
able to
distinguish the different small $x$ behavior of these two effective
gluon
distributions.

\section{Effective heavy-quark distributions in the photon}
\label{EFFECTIVE}

In this section, we compare the result of
the quark parton model (QPM) calculation of the effective
charm quark distribution
with that of the massive inhomogeneous Altarelli-Parisi(AP)
equations \cite{GHR}.
We expect that the QPM approach is appropriate at low momentum
transfer scale
where the charm quark mass effect is significant. At high momentum
transfer $Q^2/m_c^2  \gg 1$, gluon emission from charm quarks is no
longer
negligible and we need the massive-quark AP equations to sum up the
leading
effects.
On the other hand, the approach of \cite{GHR} neglects the charm
quark mass effects in gluon emission from charm quarks, and hence it
may
overestimate the gluon emission effects  at low $Q^2$.
We therefore use the QPM prescription
for the effective charm quark distribution at $Q^2 \leq 100$ GeV$^2$
and
switch to the solution of the massive quark AP equations at
$Q^2 \geq$ 100 GeV$^2$.

The QPM charm quark distribution consists of
the valence part and the sea part and is defined as
\begin{equation}
c^{\rm QPM}(x,Q^2)=c^{\rm QPM}_{v}(x,Q^2)+c^{\rm QPM}_{sea}(x,Q^2)\,,
\label{CQPM}
\end{equation}
where
\begin{mathletters}
\label{CQPMDEF}
\begin{eqnarray}
c^{\rm QPM}_{v}(x,Q^2)&=&\frac{1}{2xe_c^2}
F^\gamma_{2,c}(x,Q^2)|_{\rm direct}\,,
\label{CQPMV}\\
c^{\rm QPM}_{sea}(x,Q^2)&=&\frac{1}{2xe_c^2}
F^\gamma_{2,c}(x,Q^2)|_{\rm resolved}\,,
\label{CQPMS}
\end{eqnarray}
\end{mathletters}
with $F^\gamma_{2,c}(x,Q^2)|_{\rm direct,\,resolved}$ as given in
Eq.(\ref{CD}) and Eq.(\ref{CR}), respectively.

As is clear from these definitions, the effective charm quark
distribution
calculated by the QPM reproduces the photon structure function well,
but we should expect large process-dependent threshold corrections
when it is
used for other processes with an equivalent real charm quark in the
photon.

At high $Q^2$, we expect that emission of collinear gluons from
charm quarks becomes significant and we solve
the massive-quark inhomogeneous AP equations
\begin{mathletters}
\label{MIAPE}
\begin{eqnarray}
\frac{d q_i(x,Q^2)}{dt}&=&
  \frac{\alpha}{2\pi} e_i^2 P_{q\gamma}(x)+
  \frac{\alpha_s(Q^2)}{2\pi}\left[P_{qq}(x)\otimes q_i(x,Q^2)+
                                  P_{qg}(x)\otimes g(x,Q^2)\right]\,,
\label{MIAPEQ}\\
\frac{d g(x,Q^2)}{dt}&=&\!
\frac{\alpha_s(Q^2)}{2\pi}\!\left[P_{gq}(x)\!\otimes\!2
\left(\sum_i\!q_i(x,Q^2)\!+\!c(x,Q^2)\right)\!+\!
                                  P_{gg}(x, 4)\!\otimes\!
g(x,Q^2)\right]\,,
\label{MIAPEG}\\
\frac{d c(x,Q^2)}{dt}&=&\!
  \frac{\alpha}{2\pi} e_c^2 P_{c\gamma}(x,Q^2)\!+\!
  \frac{\alpha_s(Q^2)}{2\pi}\!\left[P_{qq}(x)\!\otimes\!
c(x,Q^2)\!+\!
                                  P_{cg}(x,Q^2)\!\otimes\!
g(x,Q^2)\right],
\label{MIAPEC}
\end{eqnarray}
\end{mathletters}
where $i=u,\,d,\,s$, and $P_{c\gamma}$ and $P_{cg}$ are the
photon and gluon to massive-quark splitting
functions, respectively, as defined by \cite{GHR}
\begin{equation}
P_{cg}(x,Q^2)=\frac{1}{2}
\theta(1/ a-x) \frac{d}{dt}\frac{w(x,m_c^2/Q^2)}{x}\,,
\quad P_{c\gamma}(x,Q^2)=6 P_{cg}(x,Q^2)\,,
\label{MSF}
\end{equation}
with $a=1+4 m_c^2/Q^2$.
The function $w(x,r)$ is defined in Eq.~(\ref{BH}).

As in the case of light quarks, it is convenient to divide the
charm-quark distribution into the valence and the sea parts:
\begin{equation}
  c(x,Q^2)=c_v(x,Q^2)+c_{sea}(x,Q^2)\,.
  \label{eq:Cbunkai}
\end{equation}
 The valence-charm-quark
comes from the photon and the sea-charm-quark comes from the gluon.
 Eq.(\ref{MIAPEC}) can then be split into the following two equations
for $c_v$ and
$c_{sea}$:
\begin{mathletters}\label{MIAPECVS}
\begin{eqnarray}
\frac{d c_v(x,Q^2)}{dt}&=&
  \frac{\alpha}{2\pi} e_c^2 P_{c\gamma}(x,Q^2)+
  \frac{\alpha_s(Q^2)}{2\pi}P_{qq}(x)\otimes c_v(x,Q^2)\,,
\label{MIAPECV}\\
\frac{d c_{sea}(x,Q^2)}{dt}&=&
  \frac{\alpha_s(Q^2)}{2\pi}\left[P_{qq}(x)\otimes c_{sea}(x,Q^2)+
                                  P_{cg}(x,Q^2)\otimes
g(x,Q^2)\right]\,,
\label{MIAPECS}
\end{eqnarray}
\end{mathletters}

The massive-quark splitting functions $P_{cg}(x,Q^2)$ and
$P_{c\gamma}(x,Q^2)$
are singular at the charm threshold $x=1/a$. Due to this singularity,
we find that
much CPU time is needed in order to get an accurate numerical
solution
when one solves Eqs.(\ref{MIAPEQ}), (\ref{MIAPEG}), (\ref{MIAPECV})
and (\ref{MIAPECS}) directly.
The numerical problem associated with the use of the massive-quark
splitting
function of \cite{GHR} is severer in the photon structure than in the
proton
structure because of the presence of the leading inhomogeneous term.
We can avoid, however,
appearance of the singular massive splitting functions, and
obtain a set of equations which contain only smooth functions
by dividing further the valence- and sea- charm-quark distributions
into the QPM part and
the remnant. The QPM part of the valence-charm-quark distribution is
defined by
Eq.(\ref{CQPMV}) and that of the sea-charm-quark distribution
is defined by Eq.(\ref{CQPMS}). The remnants
are defined through the following equations:
\begin{mathletters}\label{DCDEF}
\begin{eqnarray}
c_v(x,Q^2)&\equiv&c^{\rm QPM}_{v}(x,Q^2)+\delta c_v(x,Q^2)\,,
\label{DCV}\\
c_{sea}(x,Q^2)&\equiv&c^{\rm QPM}_{sea}(x,Q^2)+\delta
c_{sea}(x,Q^2)\,.
\label{DCS}
\end{eqnarray}
\end{mathletters}
Using Eqs.(\ref{CQPMDEF}), (\ref{MSF}),
and (\ref{MIAPECVS}), one can derive the following
equations that govern the deviations $\delta c_v$ and $\delta
c_{sea}$:
\begin{mathletters}
\label{MIAPEDC}
\begin{eqnarray}
\frac{d}{dt}\delta c_v(x,Q^2)&=&
  \frac{\alpha_s(Q^2)}{2\pi}P_{qq}(x)\otimes
    \left[c^{\rm QPM}_{v}(x,Q^2)+ \delta c_v(x,Q^2)\right]\,,
\label{MIAPEDCV}\\
\frac{d}{dt}\delta c_{sea}(x,Q^2)&=&
  \frac{\alpha_s(Q^2)}{2\pi}P_{qq}(x)\otimes
    \left[c^{\rm QPM}_{sea}(x,Q^2)+ \delta c_{sea}(x,Q^2)\right]\,,
\label{MIAPEDCS}
\end{eqnarray}
\end{mathletters}
within the similar
approximation that Gl\"uck, Hoffmann and Reya made \cite{GHR}.
Here we take the boundary conditions
\begin{equation}
\delta c_{v,sea}(x, Q_0^2)=0.
\label{eq:DCBC}
\end{equation}
Note that there now appear no massive splitting functions in these
equations.
Intuitively, this is because, the deviation of the effective charm
quark
distribution from the QPM prediction is caused by the gluon emission
from the
charm quarks, and the emission is approximated by the massless quark
splitting function $P_{qq}(x)$ in the scheme of Ref.\cite{GHR}.
A brief explanation of the numerical method which we employ
to solve these equations is given in Appendix~\ref{APXA}.

The resulting effective charm-quark distributions (multiplied by $x$)
in the photon are shown in Fig.~\ref{Fig:charm} for $Q^2$=4, 20 and
100 GeV$^2$,
where the boundary conditions (\ref{eq:DCBC}) are set at $Q_0^2=$4
GeV$^2$.
Also shown in the figure for comparison are the
valence-up-quark distributions of WHIT1-3 and WHIT4-6.
Predictions of the QPM are shown by solid lines while those of
the massive-quark inhomogeneous AP equations are shown by dash-dotted
lines.
We find that the differences
between the predictions of the two approaches
are negligibly small for the
sea-charm-quark distribution in WHIT1 to WHIT6.
{}From ~Fig.~\ref{Fig:charm}, we find that the QPM prediction for the
valence-charm-quark distribution  differs by up to about 20\%
at $Q^2=100{\rm GeV^2}$.
The shape of the prediction in the massive inhomogeneous AP equations
is softer
than the QPM one as expected.
Since the massless splitting function
$P_{qq}$ is used for describing the gluon emission from the charm
quark, the deviation from the QPM predictions in the lower $Q^2$
region
may be an overestimate.

{}From the above discussion, we conclude that the QPM calculation
of the effective charm quark distribution is appropriate
in the region $Q^2 \lsim 100{\rm GeV^2}$, given the
present experimental accuracy and theoretical uncertainties
in the charm quark mass and higher order corrections.
At sufficiently high $Q^2$,
the massless 4-flavor  AP equations should become a good
approximation. However,
we find that the charm quark threshold effect is still significant
near $x\sim 1$
even at $Q^2\sim 100$ GeV$^2$. We therefore use the massive
inhomogeneous AP
equations between $Q^2=100{\rm GeV^2}$ and
$Q^2=2500{\rm GeV^2}$ up to where
we give parametrizations of the parton distributions.
We require the continuity of all effective parton distributions at
$Q^2=$100
GeV$^2$: at 4 GeV$^2 \leq Q^2 \leq $ 100 GeV$^2$ we use
the massless $n_f=3$ inhomogeneous AP equations
with the QPM approximation to the charm quark distribution, and at
100 GeV$^2 \leq Q^2 \leq$ 2500 GeV$^2$ we use the massive
inhomogeneous
AP equations with the massive charm quark.
The boundary conditions (\ref{eq:DCBC}) are hence set at $Q_0^2=$ 100
GeV$^2$.

Finally, as is mentioned in Sec.~\ref{CHARM}, the effective bottom
quark distribution in the photon
may be approximated by the QPM calculations
($\gamma^*\gamma\rightarrow b\bar b$
and $\gamma^*g\rightarrow b\bar b$)
all the way up to $Q^2 \simeq 2500\ {\rm GeV^2}$,
and hence we do not give parametrizations of the effective b-quark
distribution in the photon.

\section{Charm production cross section}
\label{CHARMPROD}

  In this section we study the charm-quark production cross section
via the two-photon
processes by using our new effective parton distribution functions in
the photon.
  The charm quark production cross section is expected to be much
more
sensitive on the gluonic content of the photon than the photon
structure
function
\cite{DGod}.\par

\subsection{Equivalent real photon approximation}

  To calculate the charm-quark production cross section in the
two-photon
processes, we employ the equivalent real photon approximation
(EPA) to the nearly on-shell virtual photons \cite{WW,EPA}.

  In the EPA the charm-quark pair production cross section for the
process
 $e^+ e^-\rightarrow e^+ e^- c\bar c X$  is approximated by
the convolution of the effective real photon fluxes in the $e^\pm$
beam and
the cross section of the subprocess
$\gamma \gamma \rightarrow c \bar c X$:
\begin{equation}
  \sigma(e^+e^- \rightarrow e^+e^-c\bar{c}X) \cong
  \int dx_1 dx_2 D_{\gamma/e}(x_1,Q_1^2) D_{\gamma/e}(x_2,Q_2^2)
  \,\hat\sigma(\gamma_1\gamma_2\rightarrow c\bar c X)\,,
\label{eq:EPA1}
\end{equation}
where $\hat\sigma(\gamma_1\gamma_2\rightarrow c\bar c X)$
is the subprocess cross section at $W^2=sx_1x_2$, and
$D_{\gamma/e}(x_i,Q_i^2)\ \ (i=1,2)$ is the equivalent real photon
distribution in the electron ($i=1$) or in the positron($i=2$).
The c.m.~energy of the colliding $e^-e^+$ is $\sqrt{s}$, and that of
the colliding two photons is $W$. The improved form of the
photon distribution is written as \cite{HIMMZ}
\begin{equation}
  D_{\gamma/e}(x,Q^2)={\alpha \over 2\pi} {1+(1-x)^2 \over x}
\left[\log{Q^2\over t_{\rm min}}-1\right]+ {\alpha\over 2\pi} x\,, ~~
\end{equation}
where $t_{\rm min} = m_e^2x^2/(1-x)$ is the
kinematical limit of the magnitude of the lepton momentum transfer
$t$
and the second term in the right-hand-side denotes the contribution
of
the electron helicity flip amplitudes \cite{HIMMZ}.
The scale $Q^2$ should be set by  the dynamical condition that the
subprocess ``cross section'' for a virtual process is damped
at $|t| > Q^2$\cite{EPA}.

We choose  the scales $Q_i^2$ as follows. when the photon
($\gamma_i$)
couples directly to the cham quark, we set
\begin{equation}
  Q_i^2 = \min [m_c^2 + p_t^2, t_{\rm max}, t_{\rm cut}],
\label{eq:EPAQ1}
\end{equation}
where
$p_t$ is the transverse momentum of the charm quark in
the $\gamma\gamma$ c.m.~system and
$t_{\rm max}=s(1-x_i)$ is the kinematical maximum of the momentum
transfer,
and $t_{\rm cut}$ denotes the possible experimental cut on the
magnitude of the
momentum transfer.
The scale
$m_c^2+p_t^2$ can be interpreted as the virtuality of the internal
charm quark
line;
if the
virtuality of the photon is larger than that of the internal charm
quark, the production of charm quark is strongly suppressed.
If the photon $\gamma_i$ resolves into light partons that contribute
to the $c \bar c$ production subprocesses, we set
\begin{equation}
  Q_i^2 = \min [2 {\rm GeV^2}, t_{\rm max}, t_{\rm cut}],
\label{eq:EPAQ2}
\end{equation}
since the gluon content of the photon should be suppressed if its
virtuality
is much larger than the hadronic scale.
Although in principle the quark and gluon contents of a sufficiently
virtual
photon is calculable in perturbative QCD, we neglect contributions
from
virtual photons with $|t_i|>$2 GeV$^2$.
The uncertainty associated with the choice of the cut-off scale
may be estimated by changing the scale by a factor of two, and it is
at a few \% level at TRISTAN/LEP energies, while it can
be reduced significantly by reducing the $t_{\rm cut}$ value by an
anti-tagging.

To check the validity of our approximation Eq.~(\ref{eq:EPA1})
with the scale choice Eq.~(\ref{eq:EPAQ1}),
we compare the EPA prediction with the exact cross section
for the direct process $e^+e^-\rightarrow e^+e^- c \bar c$
in which both photons couple directly to the charm
quarks.
In ~Fig.~\ref{Fig:EPACa} we compare differential cross section
of charm quark pair production evaluated exactly (solid lines with
error bars)
with that of the EPA prediction (dotted lines)
against the cosine of the scattering angle $\hat\theta$ of the
subprocess
$\gamma^*\gamma^*\rightarrow c \bar c$ in the $c \bar c$ rest frame.
The exact cross section has been calculated by HELAS\cite{HELAS} with
the help
of a Monte Carlo integration package BASES\cite{basis}.

  Near the charm quark pair threshold (Fig.~\ref{Fig:EPACa}a) we find
almost exact
agreements between the EPA and the exact results.
At far above the threshold energy (Fig.~\ref{Fig:EPACa}b),
the subprocess cross section  has peaks at $\cos \hat\theta=\pm 1$
where our
EPA prediction  underestimates the cross section by $\sim$ 30\%.
{}~Fig.~\ref{Fig:EPACc} shows the charm quark pair invariant mass
($\sqrt{\hat s}$) distribution after integrating over the scattering
angle
$\hat \theta$.
Though there is a tendency that our EPA gives
slightly smaller cross section,
we find that the EPA calculation of the total cross section
agrees with the exact cross section at 1 \% level.

\subsection{Charmed  particle production cross section}

  The inclusive $\gamma\gamma\rightarrow c\bar c X$ cross section
in Eq.~(\ref{eq:EPA1}), including the resolved processes,
is described in the leading order as
\begin{eqnarray}
 d\hat\sigma(\gamma\gamma\rightarrow c\bar c X) &=&
d\hat\sigma(\gamma\gamma\rightarrow c \bar c)|_{\hat s=W^2} \nonumber
\\
&+& 2\int dx\  g(x,Q^2)\,d\hat\sigma(\gamma g \rightarrow c \bar
c)|_{\hat s=W^2x}\nonumber\\
&+& \int dx dy\ g(x,Q^2)g(y,Q^2)\,d\hat\sigma(gg\rightarrow c \bar
c)|_{\hat s=W^2xy}\nonumber\\
&+& 2\int dx dy\ \sum_{i=u,d,s}q_i(x,Q^2)\bar
q_i(y,Q^2)\,d\hat\sigma(q\bar q\rightarrow c \bar c)|_{\hat
s=W^2xy}\,,
\label{eq:EPAS}
\end{eqnarray}
where $W$ is the invariant mass of the two photons and $\sqrt{\hat
s}$ is
that of the
charm quark pair. The effective gluon
and quark distribution functions in the photon are denoted by
 $g(x,Q^2),\  q_i(x,Q^2)$, for  which we use
the parametrizations of WHIT1 to WHIT6 parton distributions.
The scale $Q^2$ of the parton distribution functions and the QCD
coupling
$\alpha_s(Q^2)$ has been chosen as follows:
$Q^2=m_c^2+p_t^2$ for $\gamma g\rightarrow c \bar c$ and
$gg\rightarrow c\bar c$,
and $Q^2=\hat s$ for $q\bar q\rightarrow c\bar c$.

We show in Fig.~\ref{Fig:EPA}  the total charmed particle
production  cross sections.
The six curves  are obtained by setting
 $m_c=1.5{\rm GeV}$, $\alpha=1/137$ and $\Lambda_4=0.4{\rm GeV}$,
and by imposing the open charm production cut $W \geq 2m_D =3.74$
GeV.
The vertical bars attached to the WHIT1, WHIT4 and WHIT6 curves
indicate  typical uncertainties in the cross section on the charm
quark mass $m_c$
between 1.3 GeV and  1.7 GeV. It is notable that uncertainty due to
the charm
quark mass is not reduced at high energies,
because the total  cross section is always dominated by the
charm-quark
pair production near the threshold. Higher order QCD
corrections\cite{DKZZ,LRSV}
also modify the cross sections.

  As can be seen from Fig.~\ref{Fig:EPA}, the total charm quark
production
cross sections at high energy $e^+ e^-$ collider experiments
are sensitive to the gluon distribution in the photon, which cannot
be
measured accurately at the current photon structure function
experiments
as we describe in Sec.~\ref{FITTOTHEDATA}.
At TRISTAN energies, the predictions using WHIT1 to WHIT3 and those
using
WHIT4 to WHIT6 are almost the same within each group. At LEP and
LEP200
energies,
the total charmed particle pair production cross section of the
two photon process
is more sensitive to the small $x$ behavior of the gluon
distribution,
and WHIT1 to WHIT6 predictions can be distinguished
provided that the uncertainties due to the charm quark mass and
higher order corrections are reduced.

Fig.~\ref{Fig:EPAa} shows the total cross sections for the process
$e^+e^-\rightarrow e^+e^-c\bar c X$, together with the
individual contribution of
the direct photon process (the first term of Eq.~(\ref{eq:EPAS})),
the once-resolved photon processes (the second term of
Eq.~(\ref{eq:EPAS})),
and the twice-resolved photon processes
(the third and the last terms of Eq.~(\ref{eq:EPAS})).
{}From Fig.~\ref{Fig:EPAa}, we find that the resolved processes,
which are
governed mainly by the gluon contents of the photon, grows much more
rapidly for WHIT6 than for WHIT1. The contribution from
the once-resolved processes overcomes that of
the direct process even at TRISTAN energies for WHIT6, while
they become comparable only at around LEP200 energy for WHIT1.
Hence the energy dependence of the once-resolved photon contribution
to
the charm quark pair production cross section will be useful to
determine
the gluon distribution.
\section{Conclusions}
\label{CONCLUSIONS}

  We have studied all the available photon structure function data
\cite{PLUTO1,PLUTO2,TASSO,JADE,TPC2G1,TPC2G2,AMY,TOPAZ,VENUS,OPAL} at
4 GeV$^2
< Q^2 <$ 100 GeV$^2$ in the leading order of perturbative QCD, and
have
found a new set of the effective scale-dependent parton distributions
in the
photon, WHIT1 to WHIT6, which are all consistent with the present
data
(Fig.~\ref{F3.1}). The six parton distributions  have systematically
different
 gluon contents (Fig.~\ref{F4.1}) and their parametrizations are
given in
Table~\ref{TBW1} to \ref{TBW6}.
  We have studied carefully the charm quark contributions to the
observed
structure functions, which are evaluated by using the lowest order
quark
parton model matrix elements ($\gamma^* \gamma \rightarrow c {\bar
c}$ and
$\gamma^* g \rightarrow c {\bar c}$), and by using the massive
inhomogeneous AP equations.
  We have found that the photon structure function has poor
sensitivity to the
gluon distribution, except at very small $x$.
In order to probe the gluon content of the photon from the
photon structure function at small $x$,
a careful analysis of experimental data is needed.

  Predictions have also been  given for the total charm quark pair
production
cross section in the two-photon collision process
including the resolved processes at $e^+ e^-$ colliders.
At PEP and PETRA energies, the difference in the predictions of WHIT1
to WHIT6
distributions is badly observable, while at TRISTAN energies WHIT4 to
WHIT6
distributions predict significantly higher cross section
than WHIT1 to WHIT3 (Fig.~\ref{Fig:EPA}).
For all of our parton distributions, WHIT1 to WHIT6,
the contribution of the once-resolved process
exceeds that of the direct process
at energies above around 200 GeV (Fig.~\ref{Fig:EPAa}).

\section*{Acknowledgment}

  We are grateful to our experimental colleagues, T.~Nozaki of AMY,
T.~Tauchi
and H.~Hayashii of TOPAZ and S.~Odaka, H.~Ohyama, T.~Oyama and
S.~Uehara of
VENUS collaborations, for keeping us informed about the latest
analysis of
their data.
  We thank M.~Drees and P.~M.~Zerwas for valuable discussions on the
charm
quark production cross section.
  We also thank J.~Kanzaki for discussions and S.~Matsumoto for
providing us with an efficient computer
program to perform the $\chi^2$ fit.
T.I. and I.W. wish to thank Japan Society for the Promotion of
Science
for the fellowship.
The work of I.W. is partially  supported by the Grant-in-Aid for
Scientific
Research from the Ministry of Education, Science and Culture of
Japan.

\appendix
\section{Three methods to solve the AP equations }
\label{APXA}

In this appendix we introduce three numerical methods for solving
the massless inhomogeneous AP equations of Eq.~(\ref{FIAPE}) and the
massive
inhomogeneous AP equations of Eq.~(\ref{MIAPE}).
  We first introduce the Mellin transformation technique in
Sec.~\ref{APX1},
and in Sec.~\ref{APX2}
we describe the recursion method that we actually use in performing
the
$\chi^2$ fit to the experimental data.
  We use the Runge-Kutta method to solve the massive  inhomogeneous
AP equations,
which is discussed in Sec.~\ref{APX3}.

  The first two methods, the Mellin transformation and the recursion
method, work
only for the massless AP equations, while the Runge-Kutta
method can be applicable to both the massive AP equations and the
massless ones.
We use all these three methods for cross checking of our numerical
results
for the massless AP equations.

\subsection{Moment method for the massless inhomogeneous AP
equations}
\label{APX1}
  The Mellin transformation $\tilde f(n)$ of $f(x)$ is defined as
\begin{equation}
        \tilde f(n)\equiv\int_0^1 dx\,x^{n-1} f(x)~~~,
\label{MELLIN}
\end{equation}
where $n$ is a complex number.
This transformation solves the convolution integrals in the
inhomogeneous
AP equations as  simple products of the Mellin transforms of
the splitting function and the parton distribution function:
\begin{equation}
\int_0^1 dx x^{n-1} P(x)\otimes q(x) = \tilde P(n) \tilde q(n)\,.
\label{eq:Motimes}
\end{equation}

The inhomogeneous AP equations with $n_f$ massless quarks
\begin{mathletters}
\label{IZUIAPE}
\begin{eqnarray}
\frac{d q_i(x,Q^2)}{dt}&=&
  \frac{\alpha}{2\pi} e_i^2 P_{q\gamma}(x)+
  \frac{\alpha_s(Q^2)}{2\pi}\left[P_{qq}(x)\otimes q_i(x,Q^2)+
                                  P_{qg}(x)\otimes
g(x,Q^2)\right]\,,\\
\frac{d g(x,Q^2)}{dt}&=&
  \!\frac{\alpha_s(Q^2)}{2\pi}\left[P_{gq}(x)\!\otimes\!
2\sum_{i=1}^{n_f}
      q_i(x,Q^2)\!+\!
                                  P_{gg}(x,n_f)\!\otimes\!
g(x,Q^2)\right]\,,
\end{eqnarray}
\end{mathletters}
with the splitting functions
\begin{mathletters}
\label{eq:mlSPLITF}
\begin{eqnarray}
        P_{qq}(z) &=& {4 \over 3}\left[ {1+z^2 \over (1-z)_+}
                      + {3 \over 2}\delta(1-z)]\right],\\
        P_{gq}(z) &=& {4 \over 3}\,{1+(1-z)^2 \over z} ,\\
        P_{qg}(z) &=& {1 \over 2}\left[z^2+(1-z)^2\right], \\
        P_{gg}(z) &=& 6\left[{z\over (1-z)_+} +{(1-z)\over z} +
z(1-z) +
          \left({11\over 12}-{n_f \over
18}\right)\delta(1-z)\right]\,,\\
        P_{q\gamma}(z)&=& 6 P_{qg}(z)\,,
\end{eqnarray}
\end{mathletters}
are transformed to the following form in the $n$-space:
\begin{mathletters}
\label{eq:MAPEq}
\begin{eqnarray}
        \frac{d\tilde q_i(n,Q^2)}{dt} &=&
        \frac{\alpha}{2\pi} e_i^2\tilde P_{q\gamma}(n) +
        \frac{\alpha_s(Q^2)}{2\pi} \left[
        \tilde P_{qq}(n) \tilde q_i(n,Q^2) + \tilde P_{qg}(n) \tilde
g(n,Q^2)\right]\,,\\
        \frac{d\tilde g(n, Q^2)}{dt} &=& \frac{\alpha_s(Q^2)}{2\pi}
\left[
                2\sum_{j=1}^{n_f} \tilde P_{gq}(n)\tilde q_j(n,Q^2) +
\tilde P_{gg}(n,n_f)
                \tilde g(n,Q^2)\right]\,,
\end{eqnarray}
\end{mathletters}
with
\begin{mathletters}
\label{eqh:Msplit}
\begin{eqnarray}
     \tilde P_{qq}(n) &=&  {8\over 3}\left[{3\over 4}-{2n+1\over
2n(n+1)}-\gamma-\Psi(n)\right],\\
      \tilde P_{gq}(n) &=& {4\over 3}\,{n^2+n+2 \over n(n^2-1)}\,,\\
      \tilde P_{qg}(n) &=& {n^2+n+2\over 2 n(n+1)(n+2)}\,,\\
      \tilde P_{gg}(n) &=& 6\left[{11\over 12} - {n_f\over
18}+{2-n\over n(n-1)}
              +{1\over (n+1)(n+2)}
                       -\gamma-\Psi(n)\right]\,,\\
      \tilde P_{q\gamma}(n)&=& 6 \tilde P_{qg}(n)\,.
\end{eqnarray}
\end{mathletters}
Here $\Psi(x)\equiv d\log\Gamma(x)/dx$ is the di-gamma function and
$\gamma=0.577\cdots$ is  Euler constant.
Eq.~(\ref{eq:MAPEq}) can be readily solved analytically
by diagonalizing with respect to the parton flavors,
and we find $\tilde q_i(n, Q^2)$ and $\tilde g(n, Q^2)$ at arbitrary
$Q^2>Q_0^2$.

The $x$-space solution, $q_i(x,Q^2)$,
can then be obtained by performing the inverse Mellin transformation
\begin{equation}
    q_i(x,Q^2) = \frac{1}{2\pi i}\int_C dn\,x^{-n}\tilde
q_i(n,Q^2),\label{eq:invM}
\end{equation}
numerically.
The complex integration path $C$ must be in the right half plane of
all
singularities of the integrands.
We choose the path \cite{GRV0}
\begin{equation}
  C=C_+-C_-,~~~
C_\pm : n=2.5 + \exp(\pm3\pi i/4)u, \ \mbox{($u$ goes from $0$ to
$\infty)$}\,.
\end{equation}
  The angle of this path, $3\pi/4$, makes the integrals converge.
For each value of $x$ in Eq.~(\ref{eq:invM}),
$|x^{-n}|=|\exp(-n\log(x))|\sim\exp(\log(x)u/\sqrt{2})$
can be neglected at sufficiently large $u$ {\it e.g.\ }\/
$u \gsim -25/\log(x) $, for slowly varying $n$-space functions,
$\tilde q_i(n,Q^2)$.

  Although the Mellin transformation method is compact and
fast, it is
not  generally useful, because  we have to restrict
our input distributions $q_i(x, Q^2_0),\  g(x, Q_0^2)$ to those
functions
whose Mellin
transformations can be analytically obtained and hence we can not use
it to solve for our sea-quark distributions (see Eq.(\ref{INPUTS})),
 and once the splitting functions
$\tilde P_{ij}(n)$ have a
certain dependence on $Q^2$, which is the case in the massive AP
equations,
we can no longer solve Eq.~(\ref{eq:MAPEq}) analytically in general.

\subsection{Recursive method for the massless  AP equations}
\label{APX2}
  The second method to solve the massless AP equations is based on
the power
expansion of the solution by the double logarithmic energy scale
parameter $s$:
\begin{equation}
  s \equiv \log (t/t_0) = \log [\,\log(Q^2/\Lambda^2) /
\log(Q_0^2/\Lambda^2)\,]\,.
\label{SDEF}
\end{equation}
  To improve the convergence of the power expansion and the behavior
at small
$x$, we introduce the rescaled
parton distribution ${\hat q}_i$, defined as
\begin{mathletters}
\begin{eqnarray}
  {\hat q}_i(x,s) \,&\equiv&\, x^2 e^{-s} q_i(x,Q^2) \,, \\
  {\hat g}(x,s) \,&\equiv&\, x^2 e^{-s} g(x,Q^2) \,.
\end{eqnarray}
\end{mathletters}
  The massless inhomogeneous AP equations for $n_f$ quark flavors,
Eq.~(\ref{FIAPE}),
can then be rewritten as follows:
\begin{mathletters}
\label{MODIAPE}
\begin{eqnarray}
  \frac{d{\hat q}_i(x,s)}{ds} &=&
       t_0 \frac{\alpha}{2 \pi} e_i^2 {\hat P}_{q\gamma}(x)
  \!+\! t_0 e^s \frac{\alpha_s(Q^2)}{2 \pi}\!
       \left[ {\hat P}_{qq}(x) \!\otimes\! {\hat q}_i(x,s)
             \!+\!{\hat P}_{qg}(x) \!\otimes\! {\hat g}(x,s) \right]
 \!-\! {\hat q}_i(x,s)\,,  \\
  \frac{d{\hat g}(x,s)}{ds} &=&
   t_0 e^s \frac{\alpha_s(Q^2)}{2 \pi} \
       \left[ 2\sum_{i=1}^{n_f}{\hat P}_{gq}(x) \otimes {\hat
q}_i(x,s) +
              {\hat P}_{gg}(x,n_f) \otimes {\hat g}(x,s) \right]
 \!-\! {\hat g}(x,s) \,,
\end{eqnarray}
\end{mathletters}
where we define the rescaled splitting functions,
\begin{equation}
   \hat P_{ij}(x)= x^2 P_{ij}(x)\,,
\end{equation}
which is regular at $ x\to 0$.
  Since the QCD running coupling $\alpha_s(Q^2)$ scales as $\exp(-s)$
$\sim 1/\log Q^2$ in the leading log approximation, the factor
$e^s \alpha_s(Q^2)$ is regarded to be a constant:
\begin{equation}
  \epsilon  =  t_0 e^s \frac{\alpha_s(Q^2)}{2 \pi}
              =  \frac{6}{33 - 2 n_F} \,,
\end{equation}
where $n_F$ is the number of light-quark flavors which governs the
running
of the
QCD coupling $\alpha_s$.
  We adopt $n_F$ =~4 for 4.0~GeV$^2$ $\leq$~$Q^2$ $\leq$~100~GeV$^2$,
and $n_F$ =~5 for $Q^2$ $>$~100~GeV$^2$, according to
Eq.~(\ref{QCDRUN}).

  Integrating Eq.~(\ref{MODIAPE}), one finds
\begin{equation}
  {\hat q}_i(x,s) =
   {\hat q}_i(x,0)
 + s t_0 \frac{\alpha}{2 \pi} e_i^2 {\hat P}_{i\gamma}(x)
 + \epsilon \sum_j
     {\hat P}_{ij}\otimes \int_0^s ds^{\prime} {\hat
q}_j(x,s^{\prime})
 - \int_0^s ds^{\prime} {\hat q}_i(x,s^{\prime}) \,,
\label{INTMODIAPE}
\end{equation}
where ${\hat q}_i(x,0)$ is nothing but the rescaled input parton
distribution at $Q^2$ =~$Q_0^2$.
  Expanding the parton distributions by powers of $s$, {\em i.e.},
\begin{equation}
  {\hat q}_i(x,s) =
    \sum_{\ell=0}^\infty \frac{s^{\ell}}{\ell !} {\hat
q}_i^{(\ell)}(x) \,,
\label{POWER}
\end{equation}
one can reduce Eq.~(\ref{INTMODIAPE}) order by order of $s$:
\begin{mathletters}
\label{REDIAPE}
\begin{eqnarray}
  {\hat q}_i^{(0)}(x) & = &
    {\hat q}_i(x,0) \,,
       \label{REDIAPE0} \\
  {\hat q}_i^{(1)}(x) & = &
    t_0 \frac{\alpha}{2 \pi} e_i^2 {\hat P}_{i\gamma}(x)
  + \epsilon \sum_j {\hat P}_{ij}(x) \otimes {\hat q}_j^{(0)}(x)
  - {\hat q}_i^{(0)}(x) \,,
       \label{REDIAPE1} \\
  {\hat q}_i^{(\ell)}(x) & = &
    \epsilon \sum_j {\hat P}_{ij}(x) \otimes {\hat q}_j^{(\ell-1)}(x)
  - {\hat q}_i^{(\ell-1)}(x)
  \qquad \qquad (\mbox{for $\ell \geq 2$}) \,.
       \label{REDIAPE2}
\end{eqnarray}
\end{mathletters}
The above equations give the input parton distribution ${\hat
q}_i(x,0)$ as
the zero-th order approximation for ${\hat q}_i(x,s)$.
 The first correction which is linear to $s$ is the sum of the
inhomogeneous
term and the terms which is driven by the zero-th approximation
(\ref{REDIAPE1}).
  The higher order corrections are determined recursively by
Eq.(\ref{REDIAPE2}).
  Summing all the contributions as in Eq.~(\ref{POWER}) up to an
appropriate order, one obtains the solution ${\hat q}_i(x,s)$ with a
given accuracy.

  This method is useful for arbitrary input parton distributions.
  This is an advantage of this method as compared to the moment
method in
Sec.~\ref{APX1}, since our input sea-quark distribution is calculated
by
the convolution integral in Eq.~(\ref{INPUTS}) and hence does not
have
analytic Mellin transform.
Furthermore, we find that the recursive method needs much less CPU
time
 than the more general Runge-Kutta method of Sec.~\ref{APX3}.
We therefore use this scheme in the actual fitting of the
experimental data.

  When one solves the homogeneous AP equations, similar method can be
used, where the scale parameter $s$ is expressed often by the QCD
coupling constant $\alpha_s(Q^2)$ as $s$
$\equiv \log [ \alpha_s(Q_0^2)/\alpha_s(Q^2) ]$.
  In this definition, one can naturally incorporate the effect of the
change
of the effective number of the quark flavors $n_F$ at the quark mass
thresholds
by simply rescaling the $s$ variable.
  However, due to the presence of the inhomogeneous term,
we cannot absorb all scales into the $s$ variable.

  Even though the recursive method is powerful for solving the
massless
inhomogeneous AP equations, it cannot be used to solve the massive AP
equations.
This is because the significant threshold effect in the massive quark
distribution and the singular mass effect in the massive splitting
functions
do not allow the power expansion like Eq.(\ref{POWER})

\subsection{Runge--Kutta method}
\label{APX3}
  The Runge-Kutta method is the general method to solve the
differential
equations and it requires a relatively large CPU time and
sufficiently large number of data points for precise calculation.
  We use this method to integrate the inhomogeneous AP equations with
a massive charm quark:
\begin{mathletters}
\label{AMAPE}
\begin{eqnarray}
\frac{d q_i(x,Q^2)}{dt}&=&
  \frac{\alpha}{2\pi} e_i^2 P_{q\gamma}(x)+
  \frac{\alpha_s(Q^2)}{2\pi}\left[P_{qq}(x)\otimes q_i(x,Q^2)+
                                  P_{qg}(x)\otimes g(x,Q^2)\right]\,,
\label{eq:17A}\\
\frac{d g(x,Q^2)}{dt}&=&
  \frac{\alpha_s(Q^2)}{2\pi}\left[\!P_{gq}(x)\!\otimes\!
2\!\left(\sum_{i=u,d,s} q_i(x,Q^2)\!+\!
c(x,Q^2)\right)\!+\!
P_{gg}(x,3)\!\otimes\!g(x,Q^2)\!\right]\!,
\label{eq:17B}\\
\frac{d c(x,Q^2)}{dt}&=&
  \frac{\alpha}{2\pi} e_c^2 P_{c\gamma}(x,Q^2)\!+\!
  \frac{\alpha_s(Q^2)}{2\pi}\left[\!P_{qq}(x)\!\otimes\!
c(x,Q^2)\!+\!
P_{cg}(x,Q^2)\!\otimes\!g(x,Q^2)\!\right]\!.
\end{eqnarray}
\end{mathletters}
After introducing the QPM component of the charm quark
distribution by Eq.~(\ref{CQPMDEF}), the
deviations from the QPM predictions as defined in  Eq.~(\ref{DCDEF})
satisfy
the following equations:
\begin{mathletters}
\label{AMAPE2}
\begin{eqnarray}
\frac{d}{dt}\delta c_v(x,Q^2)&=&
  \frac{\alpha_s(Q^2)}{2\pi}P_{qq}(x)\otimes
    \left[c^{\rm QPM}_{v}(x,Q^2)+ \delta c_v(x,Q^2)\right]\,,
\\
\frac{d}{dt}\delta c_{sea}(x,Q^2)&=&
  \frac{\alpha_s(Q^2)}{2\pi}P_{qq}(x)\otimes
    \left[c^{\rm QPM}_{sea}(x,Q^2)+ \delta c_{sea}(x,Q^2)\right]\,.
\end{eqnarray}
\end{mathletters}
It is the above two equations with the boundary condition
Eq.(\ref{eq:DCBC}) that we solve by using the Runge-Kutta method.
  It is difficult to apply the previous two methods,
the Mellin transformation method or the recursion method, to solve
the
above  equations.
For the Mellin transformation method, we can not
 analytically Mellin transform $c^{\rm QPM}_{v,sea}$.
For the recursive method of Sec.~\ref{APX2},
the expansion of  $w(z,r)$ in powers of $s$
fails at the threshold $z=1/a=1/(1+4m_c^2/Q^2)$.

  To solve the massive AP equations of Eqs.~(\ref{eq:17A}),
(\ref{eq:17B}),
(\ref{AMAPE2}) we  use
the fourth order Runge--Kutta method with adaptive step size control
for
calculating the evolution of $t$.
By discretizing the $x$ variable as $x_j\ (j=0,1\cdots,N)$,
we solve the AP equations as a set of $5(N+1)$ ordinary differential
equations, while performing the convolution integral $P(x)\otimes
q(x)$
by the Simpson integration.
We choose
\begin{mathletters}
\begin{eqnarray}
   x_j&=&\tanh\left(\sinh (-2 + 0.04 j )\right)/2+1/2,~~~(j=0,
\cdots, 110)\,,\\
   x_{111}&=&1\,,
\end{eqnarray}
\end{mathletters}
so that the data points are dense near the both ends $x\sim0$ and
$x\sim 1$.
We find that $N = 111$ is sufficient to archive an
accuracy of 1\%.

\section{ WHIT1\/--\/WHIT6 Parametrizations }
\label{Parametrization}

For practical use of the WHIT parton distributions which are given by
the
standard valence-quark input parameters of Eq.~(\ref{STANDARD}) and
the various gluon input parameters as listed in Table~\ref{TB3}, we
give
their convenient parametrizations (or simple prescriptions to
calculate them). The parametrizations
are given in three different $Q^2$ regions,
$4{\rm GeV^2}\le Q^2 \le 100{\rm GeV^2}$,
$100{\rm GeV^2}\le Q^2 \le 2500{\rm GeV^2}$, and
$Q^2 \le 4{\rm GeV^2}$. The FORTRAN code for the distributions that
we use for generating curves in this paper is obtainable from
{\tt kaoru@kekvax.kek.jp}
or from {\tt kaoru@jpnkekvx.bitnet}

\subsection{Parametrizations in the region  \protect\boldmath
            ${\bf 4}{\rm\bf GeV^2}\le Q^{\bf 2} \le {\bf 100}{\rm\bf
GeV^2}$}
In this $Q^2$ region, we parametrize those solutions of
the massless 3-flavor AP equations which are described in
Sec.~\ref{FIT}.
The heavy quark distributions are calculated by the quark parton
model
as is mentioned in Sec.~\ref{EFFECTIVE}.
Therefore, it is sufficient to give parametrizations for the
valence-quark,
the gluon and the sea-quark distributions.

The valence-quark distribution is parametrized in the
following functional form:
\begin{equation}
x q_v(x,s)/\alpha=(A_0+A_1 x+A_2 x^2)\,x^B (1-x)^C\,,
\label{PARQV}
\end{equation}
where $s$ is defined in Eq.~(\ref{SDEF}) with $Q_0^2=4{\rm GeV^2}$
and
$\Lambda=\Lambda_4=400{\rm MeV}$, and $A_i$'s, $B$ and $C$ are
polynomials of $s$
of at most 4-th order. The coefficients of these polynomials can be
found in Table~\ref{TBW1} to \ref{TBW6}.
Note that the valence-quark distributions are
common for WHIT1 to WHIT3 and for WHIT4 to WHIT6.

For the gluon distribution, we take the following form:
\begin{equation}
x g(x,s)/\alpha=A x^B (1-x)^C+(A'_{0}+A'_{1} x)\,x^{B'} (1-x)^{C'}\,.
\label{PARG}
\end{equation}
As in the case of the valence-quark, $A$'s, $B$'s, and $C$'s are
polynomials of $s$ of at most 4-th order.
The coefficients of
these polynomials are also found in Table~\ref{TBW1} to \ref{TBW6}.
Note that the second term in Eq.~(\ref{PARG}) is common for WHIT1 to
WHIT3 and for WHIT4 to WHIT6. The reason is that
it approximately represents the contribution of the gluons emitted
from
the valence-quark, and that the valence-quark distribution is common
for
WHIT1 to WHIT3 and for WHIT4 to WHIT6.

Since the input valence-quark distributions are much harder than the
input gluon distributions for the WHIT parton distributions
as seen in Eq.~(\ref{STANDARD}) and Table~\ref{TB3}, the contribution
of the common part of the gluon distributions is expected to be
significant at $x\sim 1$. In fact, we can estimate the behavior
of the common part from Eq.~(\ref{IAPEG}).  In the one gluon
emission approximation, we get the following relation for the
common gluon distribution which originates from the valence-quark:
\begin{equation}
g(x)\sim P_{gq}(x)\otimes q_v(x)\,.
\label{UG1}
\end{equation}
Assuming the behavior of the valence-quark distribution
at $x\sim 1$ as $q_v(x)\sim (1-x)^{C_v}$, Eq.(\ref{UG1}) leads to
\begin{equation}
g(x)\sim (1-x)^{C_v+1}\qquad {\rm as}\quad x\rightarrow 1\,.
\label{UG2}
\end{equation}
This common part of the gluon distributions is dominant
at $x\sim 1$ in the WHIT parton distributions, because
$C_v<1$ and $C_g\ge 3$ for all the input parton distributions.
One may find that the above crude estimation of the behavior of
the common part of the gluon distributions at $x\sim 1$ works
rather well even in quantitative sense by looking at the relevant
entries of Table~\ref{TBW1} to \ref{TBW6}.

As for the sea-quark distribution, we parametrize it
by the following functional form:
\begin{equation}
x q_{sea}(x,s)/\alpha=A\,x^{(B_0+B_1x)} (1-x)^C\,.
\label{PARQSEA}
\end{equation}
The coefficients that describe the $s$ dependence of $A$, $B$'s
and $C$ are also listed in Table~\ref{TBW1} to \ref{TBW6}.

The effective charm quark distribution is calculated by the quark
parton model
by using
Eqs.~(\ref{CQPM}) and (\ref{CQPMDEF}).
In the calculation of the sea-charm-quark distribution in
Eq.~(\ref{CQPMS}), the above parametrization of
the gluon distribution for each WHIT parton distribution is used.
An efficient integration routine for the convolution in the
calculation
of the  sea-charm-quark distribution is included in our FORTRAN code
for WHIT parametrizations.

Our parametrizations reproduce the distributions
within 5\% in the region of $0.001\le s\le 0.8$ (corresponding to
the region of $4.01{\rm GeV^2}\le Q^2\le 207{\rm GeV^2}$) and
$0.001\le x\le 0.99$ for the up, down and gluon distributions.
Note that the up and down quark distributions are written in terms of
the valence- and sea- quark distributions as shown in
Eq.(\ref{UANDD}).
In the region $0\le s\le 0.001$ (4 GeV$^2 \le Q^2 \le 4.01$ GeV$^2$),
they do not agree precisely with the distributions mainly because our
over-simplified initial distributions do not satisfy the perturbative
relations among quarks and gluon distributions such as
Eq.~(\ref{UG2}).
As for the charm quark distributions, our convolution integral
routine is
sufficiently accurate that their error is at most 5\% reflecting
the error in the parametrizations of the gluon distributions.

\subsection{Parametrizations in the region  \protect\boldmath
            ${\bf 100}{\rm\bf GeV^2}\le Q^{\bf 2} \le {\bf
2500}{\rm\bf GeV^2}$}
In this $Q^2$ region, we parametrize the solutions of the massive
inhomogeneous
AP equations obtained in Sec.~\ref{EFFECTIVE}, in which the charm
quark
mass is retained.
According to the prescription of Sec.~\ref{EFFECTIVE},
we give the deviations from the QPM predictions for the
valence-charm-quark and the  sea-charm-quark
distributions that are defined in Eq.~(\ref{DCDEF}),
in addition to the valence-light-quark,
the gluon and the sea-light-quark distributions.

The valence-light-quark, the gluon and the sea-light-quark
distributions are parametrized in exactly the same manner as in
the previous lower $Q^2$ case, except that $Q_0$ and $\Lambda$ in
the definition of $s$ are now changed to $Q_0^2=100{\rm GeV^2}$ and
$\Lambda=\Lambda_5=302.3$ MeV.

The deviation in the valence-charm-quark distribution is parametrized
as
\begin{equation}
x\,\delta c_v(x,s)/\alpha=(A_0+A_1 x+A_2 x^2+A_3 x^3)\,x^B (1-x)^C\,.
\label{PARDCV}
\end{equation}
Note that, as seen from  Eq.~(\ref{MIAPEDCV}), $\delta c_v$ is
completely universal,
{\em i.e.} it is common for all six WHIT parton distributions.
The parametrization of the deviation in the sea-charm-quark
distribution takes the following form:
\begin{equation}
x\,\delta c_{sea}(x,s)/\alpha=A\,x^{(B_0+B_1x)} (1-x)^C\,,
\label{PARDCS}
\end{equation}
which is the same form as the parametrization of
the sea-light-quark distribution. The total charm quark
distribution is given by Eqs.~(\ref{eq:Cbunkai}), ~(\ref{DCDEF}).
The QPM part of the sea-charm-quark
 distribution
is calculated by using the parametrization of the
corresponding gluon distribution.

The relative error of these parametrizations
is less than 5\% for
the up, down and gluon distributions in the region of
$0\le s\le 0.4$ (corresponding to
the region of $100{\rm GeV^2}\le Q^2\le 3120{\rm GeV^2}$),
and $0.00125\le x\le 0.99$. Note that the up and down quark
distributions are related to the valence- and the sea- light-quark
distributions by Eq.~(\ref{UANDD}).
For the charm quark distributions, the
accuracy is also within 5\% in the same $Q^2$ region but
in the different $x$ region: $0.00125\le x\le 0.99/a$, where
$x=1/a=1/(1+4 m_c^2/Q^2)$ represents the threshold for the charm
quark.

\subsection{Prescription in the region of \protect\boldmath
            $Q^{\bf 2} \le {\bf 4}{\rm\bf GeV^2}$}
Finally, we give a prescription for
the parton distributions at  $Q^2 \leq Q^2_0$ = 4 GeV$^2$.
This is because occasionally one wants to estimate the effects at
lower
$Q^2$ region. We give a very crude estimate here that can be used in
such
cases, rather than setting all the distributions to zero or freezing
the
scale-dependences.

As is explained in the main text, we give initial parton
distributions at $Q^2=Q_0^2=4{\rm GeV^2}$ and evolve them to
higher $Q^2$ by using the inhomogeneous AP equations.
Since the AP equations cannot generally be
solved in the backward direction due to its instability,
there are no ways to calculate the correct parton distributions
below $4{\rm GeV^2}$.

Our prescription to estimate the parton distributions in
the region of $Q^2 \le 4{\rm GeV}^2$ is simply to
multiply the factor $\log(Q^2/\Lambda_4^2)/\log(4 {\rm
GeV}^2/\Lambda_4^2)$
to the corresponding parton distributions
at $Q^2=4{\rm GeV^2}$ {\em i.e.} the input parton distributions.
This crude prescription
gives a reasonably good estimate for the valence quark distributions,
but
the resulting gluon distribution may not be realistic.
Therefore, one has to be careful when using this estimate
at $Q^2 < 4$  GeV$^2$.

In our application, the charm quark pair production process probes
the photon
structure down to $m_c^2$, which can take the lowest value of
$(1.3 {\rm GeV})^2 = 1.69$ GeV$^2$ for $m_c=1.3$GeV.
We have checked that the effect of modifying the prescription at
$Q^2 < 4$ GeV$^2$ is negligibly small for the total charm quark
production
cross section.

\begin{table}
\caption{The data of $F_2^\gamma$ adopted in the fit of the
valence-quark parameters. }
\label{TB1}
\begin{tabular}{llrlc}
collider & collab. & $\!\! \langle Q^2\rangle (\mbox{GeV$^2$}) \!\!$
&
  \multicolumn{1}{c}{$x$ bins} & Ref. \\
\hline
PETRA   & PLUTO   &   4.3 \quad \qquad & {\small 0.03--0.17,\
0.17--0.44,\
                                    0.44--0.80} & \cite{PLUTO1} \\
        &         &   9.2 \quad \qquad & {\small 0.06--0.23,\
0.23--0.54,\
                                    0.54--0.90} & \cite{PLUTO1} \\
        &         &  45.0 \quad \qquad & {\small 0.25--0.50,\
0.50--0.75,\
                                    0.75--0.90} & \cite{PLUTO2} \\
        & TASSO   &  23.0 \quad \qquad & {\small 0.20--0.40,\
0.40--0.60,\
                                    0.60--0.80,\ 0.80--0.98} &
\cite{TASSO} \\
        & JADE    &  24.0 \quad \qquad & {\small 0.10--0.20,\
0.20--0.40,\
                                    0.40--0.60,\ 0.60--0.90} &
\cite{JADE} \\
        &         & 100.0 \quad \qquad & {\small 0.10--0.30,\
0.30--0.60,\
                                    0.60--0.90} & \cite{JADE} \\
\hline
PEP & TPC/$2\gamma$ & 5.1 \quad \qquad & {\small 0.02--0.20,\
0.20--0.36,\
                                    0.36--0.74} & \cite{TPC2G1} \\
        &         &  20.0 \quad \qquad & {\small 0.196--0.386,\
0.386--0.611,\
                                    0.611--0.963} & \cite{TPC2G2} \\
\hline
TRISTAN & AMY     &  73.0 \quad \qquad & {\small 0.125--0.375,\
0.375--0.625,\
                                    0.625--0.875} & \cite{AMY} \\
        & TOPAZ   &   5.1 \quad \qquad & {\small 0.076--0.20} &
\cite{TOPAZ} \\
        &         &  16.0 \quad \qquad & {\small 0.15--0.33,\
0.33--0.78}
                                  & \cite{TOPAZ} \\
        &         &  80.0 \quad \qquad & {\small 0.32--0.59,\
0.59--0.98}
                                  & \cite{TOPAZ} \\
        & VENUS   &  40.0 \quad \qquad & {\small 0.09--0.27,\
0.27--0.45,\
                                    0.45--0.63,\ 0.63--0.81} &
\cite{VENUS} \\
        &         &  90.0 \quad \qquad & {\small 0.19--0.37,\
0.37--0.55,\
                                    0.55--0.73,\ 0.73--0.91} &
\cite{VENUS} \\
\hline
LEP     & OPAL    &   5.9 \quad \qquad & {\small 0.091--0.283,\
0.283--0.649}
                                  & \cite{OPAL} \\
        &         &  14.7 \quad \qquad & {\small 0.137--0.324,\
0.324--0.522,\
                                    0.522--0.836} & \cite{OPAL} \\
\end{tabular}
\end{table}
\begin{table}
\squeezetable
\caption{
  The minimal $\chi^2$ and the valence-quark parameters as obtained
by the best
  fit. Whenever available, we have taken into account correlation in
errors.
  Degree of the freedom of the fit is 47$-$3=44.}
\label{TB2}
\begin{tabular}{ccccccccc}
 \multicolumn{2}{c}{gluon} & best fit &
  \multicolumn{3}{c}{valence-quark parameters} &
  \multicolumn{3}{c}{correlations} \\
       $A_g$ & $\!\! C_g$ & $\chi^2$ & $A_v$ & $B_v$ & $C_v$ &
  $\!\! \rho \scriptstyle (A_v,B_v) \!\! $ &
  $\!\! \rho \scriptstyle (A_v,C_v) \!\! $ &
  $\!\! \rho \scriptstyle (B_v,C_v) \!\! $ \\
\hline
 0.5 & $\!\!$  3 & 51.6 & 0.930(79) & 0.50(17) & 0.24(25) &
                                   $-$0.52 & $-$0.75 & 0.88 \\
 0.5 & $\!\!$  6 & 53.2 & 0.933(78) & 0.51(17) & 0.28(26) &
                                   $-$0.52 & $-$0.75 & 0.88 \\
 0.5 & $\!\!$  9 & 54.0 & 0.938(78) & 0.49(16) & 0.28(25) &
                                   $-$0.52 & $-$0.75 & 0.88 \\
 0.5 & $\!\!$ 15 & 54.3 & 0.948(78) & 0.44(16) & 0.26(25) &
                                   $-$0.52 & $-$0.75 & 0.88 \\
\hline
 1.0 & $\!\!$  3 & 54.0 & 0.873(76) & 0.77(21) & 0.41(29) &
                                   $-$0.52 & $-$0.73 & 0.88 \\
 1.0 & $\!\!$  6 & 58.2 & 0.882(74) & 0.77(20) & 0.48(30) &
                                   $-$0.53 & $-$0.74 & 0.89 \\
 1.0 & $\!\!$  9 & 60.2 & 0.892(74) & 0.71(19) & 0.47(29) &
                                   $-$0.53 & $-$0.74 & 0.88 \\
 1.0 & $\!\!$ 15 & 60.4 & 0.911(75) & 0.61(17) & 0.42(28) &
                                   $-$0.53 & $-$0.75 & 0.88 \\
\hline
 1.5 & $\!\!$  3 & 59.8 & 0.821(73) & 1.11(26) & 0.62(34) &
                                   $-$0.53 & $-$0.72 & 0.89 \\
 1.5 & $\!\!$  6 & 67.8 & 0.837(71) & 1.05(24) & 0.70(34) &
                                   $-$0.54 & $-$0.73 & 0.89 \\
 1.5 & $\!\!$  9 & 70.9 & 0.853(71) & 0.95(22) & 0.67(33) &
                                   $-$0.54 & $-$0.73 & 0.89 \\
 1.5 & $\!\!$ 15 & 70.1 & 0.879(72) & 0.79(19) & 0.58(31) &
                                   $-$0.53 & $-$0.74 & 0.89 \\
\end{tabular}
\end{table}
\begin{table}
\squeezetable
\caption{
The $\chi^2$ values with the standard valence-quark parameters. }
\label{TB3}
\begin{tabular}{cccccccc}
 name & \multicolumn{2}{c}{gluon} &  &
 name & \multicolumn{2}{c}{gluon} &  \\
      & $A_g$ & $\!\! C_g$ & $\chi^2$ & & $A_g$ & $C_g$ & $\chi^2$ \\
\hline
          & 0.5 & $\!\!$  1 & 50.7  &          & 1.0 & $\!\!$  1 &
54.2 \\
          & 0.5 & $\!\!$  2 & 51.1  &          & 1.0 & $\!\!$  2 &
55.0 \\
{\small WHIT1} & 0.5 & $\!\!$  3 & 51.6  & {\small WHIT4} & 1.0 &
$\!\!$  3 & 56.0 \\
          & 0.5 & $\!\!$  4 & 52.2  &          & 1.0 & $\!\!$  4 &
57.0 \\
          & 0.5 & $\!\!$  5 & 52.7  &          & 1.0 & $\!\!$  5 &
57.9 \\
          & 0.5 & $\!\!$  6 & 53.2  &          & 1.0 & $\!\!$  6 &
58.6 \\
          & 0.5 & $\!\!$  7 & 53.6  &          & 1.0 & $\!\!$  7 &
59.3 \\
          & 0.5 & $\!\!$  8 & 54.0  &          & 1.0 & $\!\!$  8 &
59.8 \\
{\small WHIT2} & 0.5 & $\!\!$  9 & 54.3  & {\small WHIT5} & 1.0 &
$\!\!$  9 & 60.2 \\
          & 0.5 & $\!\!$ 10 & 54.5  &          & 1.0 & $\!\!$ 10 &
60.6 \\
          & 0.5 & $\!\!$ 11 & 54.7  &          & 1.0 & $\!\!$ 11 &
60.9 \\
          & 0.5 & $\!\!$ 12 & 54.9  &          & 1.0 & $\!\!$ 12 &
61.1 \\
          & 0.5 & $\!\!$ 13 & 55.1  &          & 1.0 & $\!\!$ 13 &
61.2 \\
          & 0.5 & $\!\!$ 14 & 55.2  &          & 1.0 & $\!\!$ 14 &
61.3 \\
{\small WHIT3} & 0.5 & $\!\!$ 15 & 55.3  & {\small WHIT6} & 1.0 &
$\!\!$ 15 & 61.4 \\
          & 0.5 & $\!\!$ 16 & 55.3  &          & 1.0 & $\!\!$ 16 &
61.5 \\
\end{tabular}
\end{table}
\textheight 245mm
\topmargin -5.0mm
\vspace*{-2.7cm}
\begin{table}
\squeezetable
\caption{Coefficients of the parametrization for WHIT1 parton
distribution in the photon. }
\label{TBW1}
\begin{tabular}{%
ccccccc
}
\multicolumn{2}{c}{
  $Q^2$}&
  \multicolumn{5}{c  }
    { $4{\rm GeV}^2\leq Q^2 \leq 100{\rm GeV}^2$}\\
\hline
\multicolumn{2}{ c }{ }&
  $s^0$ & $s^1$ & $s^2$ & $s^3$ & $s^4$ \\
\hline
 $ q_v $ &
 $ A_0 $  &
 $ 1.882 $ & $ 1.213 $ & $ 0.697 $ & $ 0 $ & $ 0 $ \\
\cline{2-7}
 &
 $ A_1 $  &
 $ 0 $ & $ -2.361 $ & $ -1.136 $ & $ 0 $ & $ 0 $ \\
\cline{2-7}
 &
 $ A_2 $  &
 $ 0 $ & $ 0.528 $ & $ 2.406 $ & $ 0 $ & $ 0 $ \\
\cline{2-7}
 &
 $ B $  &
 $ 0.500 $ & $ 0.02107 $ & $ 0.00413 $ & $ 0 $ & $ 0 $ \\
\cline{2-7}
 &
 $ C $  &
 $ 0.2500 $ & $ -0.2376 $ & $ 0.2018 $ & $ -0.0504 $ & $ 0 $ \\
\hline
 $ g $ &
 $ A $  &
 $ 2.000 $ & $ -3.28 $ & $ 2.894 $ & $ -1.561 $ & $ 0.818 $ \\
\cline{2-7}
 &
 $ B $  &
 $ 0 $ & $ -0.761 $ & $ -0.0490 $ & $ 0.446 $ & $ 0 $ \\
\cline{2-7}
 &
 $ C $  &
 $ 3.000 $ & $ 1.586 $ & $ -0.949 $ & $ 2.425 $ & $ 0 $ \\
\cline{2-7}
 &
 $ A'_{0} $  &
 $ 0 $ & $ 0.461 $ & $ 0.1041 $ & $ -0.01753 $ & $ -0.2717 $ \\
\cline{2-7}
 &
 $ A'_{1} $  &
 $ 0 $ & $ 0.00968 $ & $ -0.417 $ & $ -0.395 $ & $ 0.843 $ \\
\cline{2-7}
 &
 $ B' $  &
 $ -0.414 $ & $ -0.0606 $ & $ 0.2847 $ & $ -0.507 $ & $ 0 $ \\
\cline{2-7}
 &
 $ C' $  &
 $ 1.244 $ & $ 0.588 $ & $ -1.228 $ & $ 0.809 $ & $ 0 $ \\
\hline
 $ q_{sea} $ &
 $ A $  &
 $ 0.651 $ & $ 1.291 $ & $ -4.47 $ & $ 5.14 $ & $ -2.091 $ \\
\cline{2-7}
 &
 $ B_0 $  &
 $ -0.0382 $ & $ 0.0901 $ & $ -1.356 $ & $ 1.582 $ & $ -0.644 $ \\
\cline{2-7}
 &
 $ B_1 $  &
 $ 2.084 $ & $ 7.74 $ & $ -29.70 $ & $ 38.6 $ & $ -17.05 $ \\
\cline{2-7}
 &
 $ C $  &
 $ 7.00 $ & $ -16.08 $ & $ 46.7 $ & $ -57.1 $ & $ 23.86 $ \\
\hline
\hline
\multicolumn{2}{ c }{$Q^2$}&
  \multicolumn{5}{c }
    {$100{\rm GeV}^2\leq Q^2 \leq 2500{\rm GeV}^2$}\\
\hline
\multicolumn{2}{ c  }{ }&
  $s^0$ & $s^1$ & $s^2$ & $s^3$ & $s^4$ \\
\hline
 $ q_v $ &
 $ A_0 $  &
 $ 3.058 $ & $ 2.474 $ & $ 1.002 $ & $ 0 $ & $ 0 $ \\
\cline{2-7}
 &
 $ A_1 $  &
 $ -2.182 $ & $ -4.48 $ & $ -0.2251 $ & $ 0 $ & $ 0 $ \\
\cline{2-7}
 &
 $ A_2 $  &
 $ 1.522 $ & $ 4.31 $ & $ 1.314 $ & $ 0 $ & $ 0 $ \\
\cline{2-7}
 &
 $ B $  &
 $ 0.517 $ & $ 0.0404 $ & $ -0.02100 $ & $ 0 $ & $ 0 $ \\
\cline{2-7}
 &
 $ C $  &
 $ 0.1655 $ & $ -0.02062 $ & $ 0.0536 $ & $ 0 $ & $ 0 $ \\
\hline
 $ g $ &
 $ A $  &
 $ 0.784 $ & $ -2.238 $ & $ 16.17 $ & $ -62.5 $ & $ 83.9 $ \\
\cline{2-7}
 &
 $ B $  &
 $ -0.403 $ & $ -1.307 $ & $ 8.78 $ & $ -35.8 $ & $ 53.5 $ \\
\cline{2-7}
 &
 $ C $  &
 $ 4.45 $ & $ 1.027 $ & $ 44.6 $ & $ -160.0 $ & $ 181.6 $ \\
\cline{2-7}
 &
 $ A'_{0} $  &
 $ 0.3010 $ & $ 1.275 $ & $ -1.563 $ & $ 4.10 $ & $ -13.37 $ \\
\cline{2-7}
 &
 $ A'_{1} $  &
 $ -0.1305 $ & $ -1.245 $ & $ 2.438 $ & $ -2.539 $ & $ 12.73 $ \\
\cline{2-7}
 &
 $ B' $  &
 $ -0.489 $ & $ 0.955 $ & $ -4.40 $ & $ 10.22 $ & $ -17.13 $ \\
\cline{2-7}
 &
 $ C' $  &
 $ 1.331 $ & $ -0.2481 $ & $ 1.950 $ & $ -2.072 $ & $ 0 $ \\
\hline
 $ q_{sea} $ &
 $ A $  &
 $ 0.625 $ & $ -0.589 $ & $ 4.18 $ & $ -12.06 $ & $ 12.57 $ \\
\cline{2-7}
 &
 $ B_0 $  &
 $ -0.2492 $ & $ -0.411 $ & $ 0.966 $ & $ -2.584 $ & $ 2.670 $ \\
\cline{2-7}
 &
 $ B_1 $  &
 $ 2.100 $ & $ -5.75 $ & $ 47.8 $ & $ -140.7 $ & $ 147.6 $ \\
\cline{2-7}
 &
 $ C $  &
 $ 4.78 $ & $ 4.86 $ & $ -48.9 $ & $ 147.7 $ & $ -160.2 $ \\
\hline
 $ \delta c_v $ &
 $ A_0 $  &
 $ 0 $ & $ 0.1219 $ & $ 6.20 $ & $ -25.04 $ & $ 30.98 $ \\
\cline{2-7}
 &
 $ A_1 $  &
 $ 0 $ & $ 1.913 $ & $ -76.9 $ & $ 318. $ & $ -392. $ \\
\cline{2-7}
 &
 $ A_2 $  &
 $ 0 $ & $ -7.16 $ & $ 250.3 $ & $ -1062. $ & $ 1308. $ \\
\cline{2-7}
 &
 $ A_3 $  &
 $ 0 $ & $ 3.19 $ & $ -230.1 $ & $ 1012. $ & $ -1250. $ \\
\cline{2-7}
 &
 $ B $  &
 $ 0.499 $ & $ 3.47 $ & $ -15.26 $ & $ 19.67 $ & $ 0 $ \\
\cline{2-7}
 &
 $ C $  &
 $ 0.329 $ & $ 8.24 $ & $ -38.0 $ & $ 46.3 $ & $ 0 $ \\
\hline
 $ \delta c_{sea} $ &
 $ A $  &
 $ 0 $ & $ -0.01815 $ & $ 0.002043 $ & $ -0.00413 $ & $ 0 $ \\
\cline{2-7}
 &
 $ B_0 $  &
 $ -0.3086 $ & $ -0.2565 $ & $ 0.0984 $ & $ 0 $ & $ 0 $ \\
\cline{2-7}
 &
 $ B_1 $  &
 $ 1.376 $ & $ -0.463 $ & $ 1.232 $ & $ 0 $ & $ 0 $ \\
\cline{2-7}
 &
 $ C $  &
 $ 3.65 $ & $ 0.729 $ & $ -7.57 $ & $ 7.79 $ & $ 0 $ \\
\end{tabular}
\end{table}
\vspace*{-3.2cm}
\begin{table}
\squeezetable
\caption{Coefficients of the parametrization for WHIT2 parton
distribution in the photon. }
\label{TBW2}
\begin{tabular}{%
c  c
ccccc
 }
\multicolumn{2}{ c }{ $Q^2$}&
  \multicolumn{5}{c  }
    {$4{\rm GeV}^2\leq Q^2 \leq 100{\rm GeV}^2$}\\
\hline
\multicolumn{2}{ c }{ }&
  $s^0$ & $s^1$ & $s^2$ & $s^3$ & $s^4$ \\
\hline
 $ q_v $ &
 $ A_0 $  &
 $ 1.882 $ & $ 1.213 $ & $ 0.697 $ & $ 0 $ & $ 0 $ \\
\cline{2-7}
 &
 $ A_1 $  &
 $ 0 $ & $ -2.361 $ & $ -1.136 $ & $ 0 $ & $ 0 $ \\
\cline{2-7}
 &
 $ A_2 $  &
 $ 0 $ & $ 0.528 $ & $ 2.406 $ & $ 0 $ & $ 0 $ \\
\cline{2-7}
 &
 $ B $  &
 $ 0.500 $ & $ 0.02107 $ & $ 0.00413 $ & $ 0 $ & $ 0 $ \\
\cline{2-7}
 &
 $ C $  &
 $ 0.2500 $ & $ -0.2376 $ & $ 0.2018 $ & $ -0.0504 $ & $ 0 $ \\
\hline
 $ g $ &
 $ A $  &
 $ 5.00 $ & $ -14.99 $ & $ 26.17 $ & $ -25.30 $ & $ 10.12 $ \\
\cline{2-7}
 &
 $ B $  &
 $ 0 $ & $ -0.937 $ & $ 0.410 $ & $ 0.0339 $ & $ 0 $ \\
\cline{2-7}
 &
 $ C $  &
 $ 9.00 $ & $ 0.709 $ & $ 3.118 $ & $ -0.000582 $ & $ 0 $ \\
\cline{2-7}
 &
 $ A'_{0} $  &
 $ 0 $ & $ 0.461 $ & $ 0.1041 $ & $ -0.01753 $ & $ -0.2717 $ \\
\cline{2-7}
 &
 $ A'_{1} $  &
 $ 0 $ & $ 0.00968 $ & $ -0.417 $ & $ -0.395 $ & $ 0.843 $ \\
\cline{2-7}
 &
 $ B' $  &
 $ -0.414 $ & $ -0.0606 $ & $ 0.2847 $ & $ -0.507 $ & $ 0 $ \\
\cline{2-7}
 &
 $ C' $  &
 $ 1.244 $ & $ 0.588 $ & $ -1.228 $ & $ 0.809 $ & $ 0 $ \\
\hline
 $ q_{sea} $ &
 $ A $  &
 $ 1.237 $ & $ 3.39 $ & $ -10.75 $ & $ 12.46 $ & $ -5.58 $ \\
\cline{2-7}
 &
 $ B_0 $  &
 $ -0.0727 $ & $ 0.1748 $ & $ -1.392 $ & $ 1.711 $ & $ -0.796 $ \\
\cline{2-7}
 &
 $ B_1 $  &
 $ 4.29 $ & $ 17.87 $ & $ -58.1 $ & $ 81.9 $ & $ -41.4 $ \\
\cline{2-7}
 &
 $ C $  &
 $ 14.34 $ & $ -44.9 $ & $ 119.7 $ & $ -158.5 $ & $ 75.3 $ \\
\hline
\hline
\multicolumn{2}{ c }{$Q^2$}&
  \multicolumn{5}{c }
    {$100{\rm GeV}^2\leq Q^2 \leq 2500{\rm GeV}^2$}\\
\hline
\multicolumn{2}{ c  }{ }&
  $s^0$ & $s^1$ & $s^2$ & $s^3$ & $s^4$ \\
\hline
 $ q_v $ &
 $ A_0 $  &
 $ 3.058 $ & $ 2.474 $ & $ 1.002 $ & $ 0 $ & $ 0 $ \\
\cline{2-7}
 &
 $ A_1 $  &
 $ -2.182 $ & $ -4.48 $ & $ -0.2259 $ & $ 0 $ & $ 0 $ \\
\cline{2-7}
 &
 $ A_2 $  &
 $ 1.522 $ & $ 4.30 $ & $ 1.315 $ & $ 0 $ & $ 0 $ \\
\cline{2-7}
 &
 $ B $  &
 $ 0.517 $ & $ 0.0403 $ & $ -0.02098 $ & $ 0 $ & $ 0 $ \\
\cline{2-7}
 &
 $ C $  &
 $ 0.1655 $ & $ -0.02063 $ & $ 0.0537 $ & $ 0 $ & $ 0 $ \\
\hline
 $ g $ &
 $ A $  &
 $ 1.095 $ & $ -2.388 $ & $ 9.19 $ & $ -30.32 $ & $ 34.8 $ \\
\cline{2-7}
 &
 $ B $  &
 $ -0.441 $ & $ -0.907 $ & $ 4.68 $ & $ -18.66 $ & $ 27.17 $ \\
\cline{2-7}
 &
 $ C $  &
 $ 10.99 $ & $ 4.71 $ & $ 28.01 $ & $ -127.9 $ & $ 164.0 $ \\
\cline{2-7}
 &
 $ A'_{0} $  &
 $ 0.3010 $ & $ 1.275 $ & $ -1.563 $ & $ 4.10 $ & $ -13.37 $ \\
\cline{2-7}
 &
 $ A'_{1} $  &
 $ -0.1305 $ & $ -1.245 $ & $ 2.438 $ & $ -2.539 $ & $ 12.73 $ \\
\cline{2-7}
 &
 $ B' $  &
 $ -0.489 $ & $ 0.955 $ & $ -4.40 $ & $ 10.22 $ & $ -17.13 $ \\
\cline{2-7}
 &
 $ C' $  &
 $ 1.331 $ & $ -0.2481 $ & $ 1.950 $ & $ -2.072 $ & $ 0 $ \\
\hline
 $ q_{sea} $ &
 $ A $  &
 $ 1.287 $ & $ -2.069 $ & $ 11.57 $ & $ -35.7 $ & $ 37.4 $ \\
\cline{2-7}
 &
 $ B_0 $  &
 $ -0.2340 $ & $ -0.443 $ & $ 1.235 $ & $ -3.72 $ & $ 3.84 $ \\
\cline{2-7}
 &
 $ B_1 $  &
 $ 6.46 $ & $ -10.48 $ & $ 89.8 $ & $ -284.7 $ & $ 299.8 $ \\
\cline{2-7}
 &
 $ C $  &
 $ 5.35 $ & $ 10.11 $ & $ -133.7 $ & $ 427. $ & $ -457. $ \\
\hline
 $ \delta c_v $ &
 $ A_0 $  &
 $ 0 $ & $ 0.1219 $ & $ 6.20 $ & $ -25.04 $ & $ 30.98 $ \\
\cline{2-7}
 &
 $ A_1 $  &
 $ 0 $ & $ 1.913 $ & $ -76.9 $ & $ 318. $ & $ -392. $ \\
\cline{2-7}
 &
 $ A_2 $  &
 $ 0 $ & $ -7.16 $ & $ 250.3 $ & $ -1062. $ & $ 1308. $ \\
\cline{2-7}
 &
 $ A_3 $  &
 $ 0 $ & $ 3.19 $ & $ -230.1 $ & $ 1012. $ & $ -1250. $ \\
\cline{2-7}
 &
 $ B $  &
 $ 0.499 $ & $ 3.47 $ & $ -15.26 $ & $ 19.67 $ & $ 0 $ \\
\cline{2-7}
 &
 $ C $  &
 $ 0.329 $ & $ 8.24 $ & $ -38.0 $ & $ 46.3 $ & $ 0 $ \\
\hline
 $ \delta c_{sea} $ &
 $ A $  &
 $ 0 $ & $ -0.02786 $ & $ 0.0349 $ & $ -0.02223 $ & $ 0 $ \\
\cline{2-7}
 &
 $ B_0 $  &
 $ -0.3141 $ & $ -0.425 $ & $ 0.1564 $ & $ 0 $ & $ 0 $ \\
\cline{2-7}
 &
 $ B_1 $  &
 $ 4.72 $ & $ -5.48 $ & $ 2.686 $ & $ 0 $ & $ 0 $ \\
\cline{2-7}
 &
 $ C $  &
 $ 2.961 $ & $ 0.776 $ & $ -8.28 $ & $ 9.78 $ & $ 0 $ \\
\end{tabular}
\end{table}

\vspace*{-3.2cm}
\begin{table}
\squeezetable
\caption{Coefficients of the parametrization for WHIT3 parton
distribution in the photon. }
\label{TBW3}
\begin{tabular}{%
ccccccc
 }
\multicolumn{2}{ c }{$Q^2$}&
  \multicolumn{5}{c  }
    {$4{\rm GeV}^2\leq Q^2 \leq 100{\rm GeV}^2$}\\
\hline
\multicolumn{2}{ c }{ }&
  $s^0$ & $s^1$ & $s^2$ & $s^3$ & $s^4$ \\
\hline
 $ q_v $ &
 $ A_0 $  &
 $ 1.882 $ & $ 1.213 $ & $ 0.697 $ & $ 0 $ & $ 0 $ \\
\cline{2-7}
 &
 $ A_1 $  &
 $ 0 $ & $ -2.361 $ & $ -1.136 $ & $ 0 $ & $ 0 $ \\
\cline{2-7}
 &
 $ A_2 $  &
 $ 0 $ & $ 0.528 $ & $ 2.406 $ & $ 0 $ & $ 0 $ \\
\cline{2-7}
 &
 $ B $  &
 $ 0.500 $ & $ 0.02107 $ & $ 0.00413 $ & $ 0 $ & $ 0 $ \\
\cline{2-7}
 &
 $ C $  &
 $ 0.2500 $ & $ -0.2376 $ & $ 0.2018 $ & $ -0.0504 $ & $ 0 $ \\
\hline
 $ g $ &
 $ A $  &
 $ 8.00 $ & $ -28.64 $ & $ 55.9 $ & $ -57.6 $ & $ 23.66 $ \\
\cline{2-7}
 &
 $ B $  &
 $ 0 $ & $ -0.987 $ & $ 0.510 $ & $ -0.0667 $ & $ 0 $ \\
\cline{2-7}
 &
 $ C $  &
 $ 15.00 $ & $ 0.331 $ & $ 3.50 $ & $ 0.892 $ & $ 0 $ \\
\cline{2-7}
 &
 $ A'_{0} $  &
 $ 0 $ & $ 0.461 $ & $ 0.1041 $ & $ -0.01753 $ & $ -0.2717 $ \\
\cline{2-7}
 &
 $ A'_{1} $  &
 $ 0 $ & $ 0.00968 $ & $ -0.417 $ & $ -0.395 $ & $ 0.843 $ \\
\cline{2-7}
 &
 $ B' $  &
 $ -0.414 $ & $ -0.0606 $ & $ 0.2847 $ & $ -0.507 $ & $ 0 $ \\
\cline{2-7}
 &
 $ C' $  &
 $ 1.244 $ & $ 0.588 $ & $ -1.228 $ & $ 0.809 $ & $ 0 $ \\
\hline
 $ q_{sea} $ &
 $ A $  &
 $ 1.587 $ & $ 5.05 $ & $ -11.26 $ & $ 7.56 $ & $ -1.471 $ \\
\cline{2-7}
 &
 $ B_0 $  &
 $ -0.1006 $ & $ 0.2259 $ & $ -1.195 $ & $ 1.175 $ & $ -0.446 $ \\
\cline{2-7}
 &
 $ B_1 $  &
 $ 5.73 $ & $ 25.64 $ & $ -58.7 $ & $ 63.2 $ & $ -25.77 $ \\
\cline{2-7}
 &
 $ C $  &
 $ 21.36 $ & $ -72.9 $ & $ 153.2 $ & $ -167.9 $ & $ 67.4 $ \\
\hline
\hline
\multicolumn{2}{ c }{$Q^2$}&
  \multicolumn{5}{c }
    {$100{\rm GeV}^2\leq Q^2 \leq 2500{\rm GeV}^2$}\\
\hline
\multicolumn{2}{ c  }{ }&
  $s^0$ & $s^1$ & $s^2$ & $s^3$ & $s^4$ \\
\hline
 $ q_v $ &
 $ A_0 $  &
 $ 3.058 $ & $ 2.474 $ & $ 1.002 $ & $ 0 $ & $ 0 $ \\
\cline{2-7}
 &
 $ A_1 $  &
 $ -2.182 $ & $ -4.48 $ & $ -0.2264 $ & $ 0 $ & $ 0 $ \\
\cline{2-7}
 &
 $ A_2 $  &
 $ 1.522 $ & $ 4.30 $ & $ 1.315 $ & $ 0 $ & $ 0 $ \\
\cline{2-7}
 &
 $ B $  &
 $ 0.517 $ & $ 0.0403 $ & $ -0.02097 $ & $ 0 $ & $ 0 $ \\
\cline{2-7}
 &
 $ C $  &
 $ 0.1655 $ & $ -0.02064 $ & $ 0.0537 $ & $ 0 $ & $ 0 $ \\
\hline
 $ g $ &
 $ A $  &
 $ 1.270 $ & $ -2.817 $ & $ 5.74 $ & $ -13.27 $ & $ 12.68 $ \\
\cline{2-7}
 &
 $ B $  &
 $ -0.461 $ & $ -0.817 $ & $ 3.32 $ & $ -12.96 $ & $ 18.93 $ \\
\cline{2-7}
 &
 $ C $  &
 $ 17.21 $ & $ 1.257 $ & $ 50.5 $ & $ -276.1 $ & $ 490. $ \\
\cline{2-7}
 &
 $ A'_{0} $  &
 $ 0.3010 $ & $ 1.275 $ & $ -1.563 $ & $ 4.10 $ & $ -13.37 $ \\
\cline{2-7}
 &
 $ A'_{1} $  &
 $ -0.1305 $ & $ -1.245 $ & $ 2.438 $ & $ -2.539 $ & $ 12.73 $ \\
\cline{2-7}
 &
 $ B' $  &
 $ -0.489 $ & $ 0.955 $ & $ -4.40 $ & $ 10.22 $ & $ -17.13 $ \\
\cline{2-7}
 &
 $ C' $  &
 $ 1.331 $ & $ -0.2481 $ & $ 1.950 $ & $ -2.072 $ & $ 0 $ \\
\hline
 $ q_{sea} $ &
 $ A $  &
 $ 1.850 $ & $ -3.67 $ & $ 27.14 $ & $ -106.6 $ & $ 130.9 $ \\
\cline{2-7}
 &
 $ B_0 $  &
 $ -0.2299 $ & $ -0.497 $ & $ 2.464 $ & $ -9.95 $ & $ 12.32 $ \\
\cline{2-7}
 &
 $ B_1 $  &
 $ 10.42 $ & $ -10.74 $ & $ 132.7 $ & $ -539. $ & $ 656. $ \\
\cline{2-7}
 &
 $ C $  &
 $ 4.07 $ & $ 4.11 $ & $ -171.9 $ & $ 707. $ & $ -859. $ \\
\hline
 $ \delta c_v $ &
 $ A_0 $  &
 $ 0 $ & $ 0.1219 $ & $ 6.20 $ & $ -25.04 $ & $ 30.98 $ \\
\cline{2-7}
 &
 $ A_1 $  &
 $ 0 $ & $ 1.913 $ & $ -76.9 $ & $ 318. $ & $ -392. $ \\
\cline{2-7}
 &
 $ A_2 $  &
 $ 0 $ & $ -7.16 $ & $ 250.3 $ & $ -1062. $ & $ 1308. $ \\
\cline{2-7}
 &
 $ A_3 $  &
 $ 0 $ & $ 3.19 $ & $ -230.1 $ & $ 1012. $ & $ -1250. $ \\
\cline{2-7}
 &
 $ B $  &
 $ 0.499 $ & $ 3.47 $ & $ -15.26 $ & $ 19.67 $ & $ 0 $ \\
\cline{2-7}
 &
 $ C $  &
 $ 0.329 $ & $ 8.24 $ & $ -38.0 $ & $ 46.3 $ & $ 0 $ \\
\hline
 $ \delta c_{sea} $ &
 $ A $  &
 $ 0 $ & $ -0.01948 $ & $ 0.02861 $ & $ -0.02036 $ & $ 0 $ \\
\cline{2-7}
 &
 $ B_0 $  &
 $ -0.413 $ & $ -0.439 $ & $ 0.1810 $ & $ 0 $ & $ 0 $ \\
\cline{2-7}
 &
 $ B_1 $  &
 $ 5.19 $ & $ -7.40 $ & $ 3.40 $ & $ 0 $ & $ 0 $ \\
\cline{2-7}
 &
 $ C $  &
 $ 2.359 $ & $ 0.977 $ & $ -7.73 $ & $ 9.48 $ & $ 0 $ \\
\end{tabular}
\end{table}
\vspace*{-3.2cm}
\begin{table}
\squeezetable
\caption{Coefficients of the parametrization for WHIT4 parton
distribution in the photon. }
\label{TBW4}
\begin{tabular}{%
c  c
ccccc
 }
\multicolumn{2}{ c }{ $Q^2$}&
  \multicolumn{5}{c  }
    { $4{\rm GeV}^2\leq Q^2 \leq 100{\rm GeV}^2$}\\
\hline
\multicolumn{2}{ c }{ }&
  $s^0$ & $s^1$ & $s^2$ & $s^3$ & $s^4$ \\
\hline
 $ q_v $ &
 $ A_0 $  &
 $ 2.540 $ & $ 2.000 $ & $ 0.718 $ & $ 0 $ & $ 0 $ \\
\cline{2-7}
 &
 $ A_1 $  &
 $ 0.0623 $ & $ -7.01 $ & $ 0.1251 $ & $ 0 $ & $ 0 $ \\
\cline{2-7}
 &
 $ A_2 $  &
 $ -0.1642 $ & $ -0.436 $ & $ 10.48 $ & $ -5.20 $ & $ 0 $ \\
\cline{2-7}
 &
 $ B $  &
 $ 0.699 $ & $ -0.02796 $ & $ -0.00365 $ & $ 0 $ & $ 0 $ \\
\cline{2-7}
 &
 $ C $  &
 $ 0.442 $ & $ -1.255 $ & $ 1.941 $ & $ -0.995 $ & $ 0 $ \\
\hline
 $ g $ &
 $ A $  &
 $ 4.00 $ & $ -9.40 $ & $ 15.55 $ & $ -14.50 $ & $ 5.47 $ \\
\cline{2-7}
 &
 $ B $  &
 $ 0 $ & $ -1.142 $ & $ 1.034 $ & $ -0.441 $ & $ 0 $ \\
\cline{2-7}
 &
 $ C $  &
 $ 3.000 $ & $ 0.872 $ & $ 1.006 $ & $ 0.356 $ & $ 0 $ \\
\cline{2-7}
 &
 $ A'_{0} $  &
 $ 0 $ & $ 0.602 $ & $ 0.509 $ & $ -2.054 $ & $ 1.392 $ \\
\cline{2-7}
 &
 $ A'_{1} $  &
 $ 0 $ & $ -0.0922 $ & $ -1.899 $ & $ 4.18 $ & $ -2.494 $ \\
\cline{2-7}
 &
 $ B' $  &
 $ -0.2895 $ & $ 0.376 $ & $ -1.719 $ & $ 1.116 $ & $ 0 $ \\
\cline{2-7}
 &
 $ C' $  &
 $ 1.439 $ & $ -0.557 $ & $ 0.366 $ & $ 0.733 $ & $ -0.762 $ \\
\hline
 $ q_{sea} $ &
 $ A $  &
 $ 1.308 $ & $ 2.315 $ & $ -7.88 $ & $ 8.26 $ & $ -3.004 $ \\
\cline{2-7}
 &
 $ B_0 $  &
 $ -0.0373 $ & $ 0.0563 $ & $ -1.133 $ & $ 1.185 $ & $ -0.418 $ \\
\cline{2-7}
 &
 $ B_1 $  &
 $ 2.103 $ & $ 4.85 $ & $ -17.81 $ & $ 20.62 $ & $ -7.94 $ \\
\cline{2-7}
 &
 $ C $  &
 $ 7.00 $ & $ -10.17 $ & $ 26.00 $ & $ -29.60 $ & $ 12.27 $ \\
\hline
\hline
\multicolumn{2}{ c }{ $Q^2$}&
  \multicolumn{5}{c }
    { $100{\rm GeV}^2\leq Q^2 \leq 2500{\rm GeV}^2$}\\
\hline
\multicolumn{2}{ c  }{ }&
  $s^0$ & $s^1$ & $s^2$ & $s^3$ & $s^4$ \\
\hline
 $ q_v $ &
 $ A_0 $  &
 $ 4.27 $ & $ 3.096 $ & $ 1.619 $ & $ 0 $ & $ 0 $ \\
\cline{2-7}
 &
 $ A_1 $  &
 $ -4.74 $ & $ -6.90 $ & $ -2.430 $ & $ 0 $ & $ 0 $ \\
\cline{2-7}
 &
 $ A_2 $  &
 $ 2.837 $ & $ 6.47 $ & $ 4.09 $ & $ 0 $ & $ 0 $ \\
\cline{2-7}
 &
 $ B $  &
 $ 0.678 $ & $ -0.0394 $ & $ 0.01756 $ & $ 0 $ & $ 0 $ \\
\cline{2-7}
 &
 $ C $  &
 $ 0.1728 $ & $ -0.02479 $ & $ 0.1446 $ & $ 0 $ & $ 0 $ \\
\hline
 $ g $ &
 $ A $  &
 $ 1.384 $ & $ -2.455 $ & $ 8.94 $ & $ -29.06 $ & $ 37.1 $ \\
\cline{2-7}
 &
 $ B $  &
 $ -0.442 $ & $ -0.719 $ & $ 2.961 $ & $ -12.09 $ & $ 19.16 $ \\
\cline{2-7}
 &
 $ C $  &
 $ 4.21 $ & $ 2.524 $ & $ 10.03 $ & $ -18.27 $ & $ 2.162 $ \\
\cline{2-7}
 &
 $ A'_{0} $  &
 $ 0.2992 $ & $ 1.179 $ & $ -1.915 $ & $ 7.26 $ & $ -18.39 $ \\
\cline{2-7}
 &
 $ A'_{1} $  &
 $ -0.1600 $ & $ -1.114 $ & $ 2.939 $ & $ -6.66 $ & $ 19.23 $ \\
\cline{2-7}
 &
 $ B' $  &
 $ -0.483 $ & $ 0.755 $ & $ -3.80 $ & $ 10.75 $ & $ -19.93 $ \\
\cline{2-7}
 &
 $ C' $  &
 $ 1.297 $ & $ -0.1669 $ & $ 1.906 $ & $ -2.057 $ & $ 0 $ \\
\hline
 $ q_{sea} $ &
 $ A $  &
 $ 1.188 $ & $ -1.396 $ & $ 8.71 $ & $ -25.42 $ & $ 24.92 $ \\
\cline{2-7}
 &
 $ B_0 $  &
 $ -0.2448 $ & $ -0.419 $ & $ 1.007 $ & $ -2.689 $ & $ 2.517 $ \\
\cline{2-7}
 &
 $ B_1 $  &
 $ 1.942 $ & $ -6.04 $ & $ 50.3 $ & $ -147.8 $ & $ 148.1 $ \\
\cline{2-7}
 &
 $ C $  &
 $ 5.42 $ & $ 6.11 $ & $ -53.8 $ & $ 163.2 $ & $ -171.6 $ \\
\hline
 $ \delta c_v $ &
 $ A_0 $  &
 $ 0 $ & $ 0.1219 $ & $ 6.20 $ & $ -25.04 $ & $ 30.98 $ \\
\cline{2-7}
 &
 $ A_1 $  &
 $ 0 $ & $ 1.913 $ & $ -76.9 $ & $ 318. $ & $ -392. $ \\
\cline{2-7}
 &
 $ A_2 $  &
 $ 0 $ & $ -7.16 $ & $ 250.3 $ & $ -1062. $ & $ 1308. $ \\
\cline{2-7}
 &
 $ A_3 $  &
 $ 0 $ & $ 3.19 $ & $ -230.1 $ & $ 1012. $ & $ -1250. $ \\
\cline{2-7}
 &
 $ B $  &
 $ 0.499 $ & $ 3.47 $ & $ -15.26 $ & $ 19.67 $ & $ 0 $ \\
\cline{2-7}
 &
 $ C $  &
 $ 0.329 $ & $ 8.24 $ & $ -38.0 $ & $ 46.3 $ & $ 0 $ \\
\hline
 $ \delta c_{sea} $ &
 $ A $  &
 $ 0 $ & $ -0.02821 $ & $ -0.0002649 $ & $ 0.00704 $ & $ 0 $ \\
\cline{2-7}
 &
 $ B_0 $  &
 $ -0.327 $ & $ -0.2298 $ & $ 0.0350 $ & $ 0 $ & $ 0 $ \\
\cline{2-7}
 &
 $ B_1 $  &
 $ 1.254 $ & $ 0.878 $ & $ 0.2086 $ & $ 0 $ & $ 0 $ \\
\cline{2-7}
 &
 $ C $  &
 $ 4.17 $ & $ 0.640 $ & $ -7.63 $ & $ 7.17 $ & $ 0 $ \\
\end{tabular}
\end{table}
\vspace*{-3.2cm}
\begin{table}
\squeezetable
\caption{Coefficients of the parametrization for WHIT5 parton
distribution in the photon. }
\label{TBW5}
\begin{tabular}{%
c  c
ccccc
 }
\multicolumn{2}{ c }{ $Q^2$}&
  \multicolumn{5}{c  }
    {$4{\rm GeV}^2\leq Q^2 \leq 100{\rm GeV}^2$}\\
\hline
\multicolumn{2}{ c }{ }&
  $s^0$ & $s^1$ & $s^2$ & $s^3$ & $s^4$ \\
\hline
 $ q_v $ &
 $ A_0 $  &
 $ 2.540 $ & $ 2.000 $ & $ 0.718 $ & $ 0 $ & $ 0 $ \\
\cline{2-7}
 &
 $ A_1 $  &
 $ 0.0623 $ & $ -7.01 $ & $ 0.1251 $ & $ 0 $ & $ 0 $ \\
\cline{2-7}
 &
 $ A_2 $  &
 $ -0.1642 $ & $ -0.436 $ & $ 10.48 $ & $ -5.20 $ & $ 0 $ \\
\cline{2-7}
 &
 $ B $  &
 $ 0.699 $ & $ -0.02796 $ & $ -0.00365 $ & $ 0 $ & $ 0 $ \\
\cline{2-7}
 &
 $ C $  &
 $ 0.442 $ & $ -1.255 $ & $ 1.941 $ & $ -0.995 $ & $ 0 $ \\
\hline
 $ g $ &
 $ A $  &
 $ 10.00 $ & $ -34.0 $ & $ 69.0 $ & $ -75.3 $ & $ 32.3 $ \\
\cline{2-7}
 &
 $ B $  &
 $ 0 $ & $ -1.126 $ & $ 0.926 $ & $ -0.393 $ & $ 0 $ \\
\cline{2-7}
 &
 $ C $  &
 $ 9.00 $ & $ 0.481 $ & $ 3.20 $ & $ -0.347 $ & $ 0 $ \\
\cline{2-7}
 &
 $ A'_{0} $  &
 $ 0 $ & $ 0.602 $ & $ 0.509 $ & $ -2.054 $ & $ 1.392 $ \\
\cline{2-7}
 &
 $ A'_{1} $  &
 $ 0 $ & $ -0.0922 $ & $ -1.899 $ & $ 4.18 $ & $ -2.494 $ \\
\cline{2-7}
 &
 $ B' $  &
 $ -0.2895 $ & $ 0.376 $ & $ -1.719 $ & $ 1.116 $ & $ 0 $ \\
\cline{2-7}
 &
 $ C' $  &
 $ 1.439 $ & $ -0.557 $ & $ 0.366 $ & $ 0.733 $ & $ -0.762 $ \\
\hline
 $ q_{sea} $ &
 $ A $  &
 $ 2.227 $ & $ 5.72 $ & $ -12.95 $ & $ 7.22 $ & $ -0.2514 $ \\
\cline{2-7}
 &
 $ B_0 $  &
 $ -0.0881 $ & $ 0.1465 $ & $ -0.975 $ & $ 0.782 $ & $ -0.2074 $ \\
\cline{2-7}
 &
 $ B_1 $  &
 $ 3.37 $ & $ 14.16 $ & $ -31.50 $ & $ 27.89 $ & $ -8.71 $ \\
\cline{2-7}
 &
 $ C $  &
 $ 15.81 $ & $ -36.3 $ & $ 77.1 $ & $ -78.1 $ & $ 29.48 $ \\
\hline
\hline
\multicolumn{2}{ c }{$Q^2$}&
  \multicolumn{5}{c }
    { $100{\rm GeV}^2\leq Q^2 \leq 2500{\rm GeV}^2$}\\
\hline
\multicolumn{2}{ c  }{ }&
  $s^0$ & $s^1$ & $s^2$ & $s^3$ & $s^4$ \\
\hline
 $ q_v $ &
 $ A_0 $  &
 $ 4.27 $ & $ 3.096 $ & $ 1.617 $ & $ 0 $ & $ 0 $ \\
\cline{2-7}
 &
 $ A_1 $  &
 $ -4.74 $ & $ -6.90 $ & $ -2.417 $ & $ 0 $ & $ 0 $ \\
\cline{2-7}
 &
 $ A_2 $  &
 $ 2.837 $ & $ 6.47 $ & $ 4.07 $ & $ 0 $ & $ 0 $ \\
\cline{2-7}
 &
 $ B $  &
 $ 0.678 $ & $ -0.0394 $ & $ 0.01750 $ & $ 0 $ & $ 0 $ \\
\cline{2-7}
 &
 $ C $  &
 $ 0.1728 $ & $ -0.02457 $ & $ 0.1440 $ & $ 0 $ & $ 0 $ \\
\hline
 $ g $ &
 $ A $  &
 $ 1.995 $ & $ -3.26 $ & $ 1.818 $ & $ 1.711 $ & $ -4.99 $ \\
\cline{2-7}
 &
 $ B $  &
 $ -0.466 $ & $ -0.610 $ & $ 1.691 $ & $ -6.68 $ & $ 10.19 $ \\
\cline{2-7}
 &
 $ C $  &
 $ 10.75 $ & $ 5.42 $ & $ 6.55 $ & $ -22.97 $ & $ 18.67 $ \\
\cline{2-7}
 &
 $ A'_{0} $  &
 $ 0.2992 $ & $ 1.179 $ & $ -1.915 $ & $ 7.26 $ & $ -18.39 $ \\
\cline{2-7}
 &
 $ A'_{1} $  &
 $ -0.1600 $ & $ -1.114 $ & $ 2.939 $ & $ -6.66 $ & $ 19.23 $ \\
\cline{2-7}
 &
 $ B' $  &
 $ -0.483 $ & $ 0.755 $ & $ -3.80 $ & $ 10.75 $ & $ -19.93 $ \\
\cline{2-7}
 &
 $ C' $  &
 $ 1.297 $ & $ -0.1669 $ & $ 1.906 $ & $ -2.057 $ & $ 0 $ \\
\hline
 $ q_{sea} $ &
 $ A $  &
 $ 2.318 $ & $ -3.76 $ & $ 20.26 $ & $ -59.5 $ & $ 59.0 $ \\
\cline{2-7}
 &
 $ B_0 $  &
 $ -0.2425 $ & $ -0.436 $ & $ 1.241 $ & $ -3.51 $ & $ 3.36 $ \\
\cline{2-7}
 &
 $ B_1 $  &
 $ 5.33 $ & $ -8.68 $ & $ 74.2 $ & $ -207.0 $ & $ 196.7 $ \\
\cline{2-7}
 &
 $ C $  &
 $ 8.48 $ & $ 9.31 $ & $ -104.1 $ & $ 280.1 $ & $ -266.3 $ \\
\hline
 $ \delta c_v $ &
 $ A_0 $  &
 $ 0 $ & $ 0.1219 $ & $ 6.20 $ & $ -25.04 $ & $ 30.98 $ \\
\cline{2-7}
 &
 $ A_1 $  &
 $ 0 $ & $ 1.913 $ & $ -76.9 $ & $ 318. $ & $ -392. $ \\
\cline{2-7}
 &
 $ A_2 $  &
 $ 0 $ & $ -7.16 $ & $ 250.3 $ & $ -1062. $ & $ 1308. $ \\
\cline{2-7}
 &
 $ A_3 $  &
 $ 0 $ & $ 3.19 $ & $ -230.1 $ & $ 1012. $ & $ -1250. $ \\
\cline{2-7}
 &
 $ B $  &
 $ 0.499 $ & $ 3.47 $ & $ -15.26 $ & $ 19.67 $ & $ 0 $ \\
\cline{2-7}
 &
 $ C $  &
 $ 0.329 $ & $ 8.24 $ & $ -38.0 $ & $ 46.3 $ & $ 0 $ \\
\hline
 $ \delta c_{sea} $ &
 $ A $  &
 $ 0 $ & $ -0.0658 $ & $ 0.1059 $ & $ -0.0663 $ & $ 0 $ \\
\cline{2-7}
 &
 $ B_0 $  &
 $ -0.2750 $ & $ -0.476 $ & $ 0.1191 $ & $ 0 $ & $ 0 $ \\
\cline{2-7}
 &
 $ B_1 $  &
 $ 6.37 $ & $ -5.32 $ & $ 1.986 $ & $ 0 $ & $ 0 $ \\
\cline{2-7}
 &
 $ C $  &
 $ 3.40 $ & $ 0.375 $ & $ -8.79 $ & $ 10.01 $ & $ 0 $ \\
\end{tabular}
\end{table}
\vspace*{-3.2cm}
\begin{table}
\squeezetable
\caption{Coefficients of the parametrization for WHIT6 parton
distribution in the photon. }
\label{TBW6}
\begin{tabular}{%
c  c
ccccc
 }
\multicolumn{2}{ c }{ $Q^2$}&
  \multicolumn{5}{c  }
    { $4{\rm GeV}^2\leq Q^2 \leq 100{\rm GeV}^2$}\\
\hline
\multicolumn{2}{ c }{ }&
  $s^0$ & $s^1$ & $s^2$ & $s^3$ & $s^4$ \\
\hline
 $ q_v $ &
 $ A_0 $  &
 $ 2.540 $ & $ 2.000 $ & $ 0.718 $ & $ 0 $ & $ 0 $ \\
\cline{2-7}
 &
 $ A_1 $  &
 $ 0.0623 $ & $ -7.01 $ & $ 0.1251 $ & $ 0 $ & $ 0 $ \\
\cline{2-7}
 &
 $ A_2 $  &
 $ -0.1642 $ & $ -0.436 $ & $ 10.48 $ & $ -5.20 $ & $ 0 $ \\
\cline{2-7}
 &
 $ B $  &
 $ 0.699 $ & $ -0.02796 $ & $ -0.00365 $ & $ 0 $ & $ 0 $ \\
\cline{2-7}
 &
 $ C $  &
 $ 0.442 $ & $ -1.255 $ & $ 1.941 $ & $ -0.995 $ & $ 0 $ \\
\hline
 $ g $ &
 $ A $  &
 $ 16.00 $ & $ -61.0 $ & $ 127.8 $ & $ -139.9 $ & $ 59.9 $ \\
\cline{2-7}
 &
 $ B $  &
 $ 0 $ & $ -1.109 $ & $ 0.845 $ & $ -0.351 $ & $ 0 $ \\
\cline{2-7}
 &
 $ C $  &
 $ 15.00 $ & $ 0.1596 $ & $ 4.18 $ & $ -0.1765 $ & $ 0 $ \\
\cline{2-7}
 &
 $ A'_{0} $  &
 $ 0 $ & $ 0.602 $ & $ 0.509 $ & $ -2.054 $ & $ 1.392 $ \\
\cline{2-7}
 &
 $ A'_{1} $  &
 $ 0 $ & $ -0.0922 $ & $ -1.899 $ & $ 4.18 $ & $ -2.494 $ \\
\cline{2-7}
 &
 $ B' $  &
 $ -0.2895 $ & $ 0.376 $ & $ -1.719 $ & $ 1.116 $ & $ 0 $ \\
\cline{2-7}
 &
 $ C' $  &
 $ 1.439 $ & $ -0.557 $ & $ 0.366 $ & $ 0.733 $ & $ -0.762 $ \\
\hline
 $ q_{sea} $ &
 $ A $  &
 $ 3.18 $ & $ 8.69 $ & $ -22.87 $ & $ 18.96 $ & $ -5.14 $ \\
\cline{2-7}
 &
 $ B_0 $  &
 $ -0.1003 $ & $ 0.1603 $ & $ -1.037 $ & $ 0.944 $ & $ -0.2915 $ \\
\cline{2-7}
 &
 $ B_1 $  &
 $ 5.69 $ & $ 18.67 $ & $ -46.7 $ & $ 50.5 $ & $ -18.35 $ \\
\cline{2-7}
 &
 $ C $  &
 $ 21.49 $ & $ -56.5 $ & $ 129.3 $ & $ -145.9 $ & $ 57.5 $ \\
\hline
\hline
\multicolumn{2}{ c }{$Q^2$}&
  \multicolumn{5}{c }
    {$100{\rm GeV}^2\leq Q^2 \leq 2500{\rm GeV}^2$}\\
\hline
\multicolumn{2}{ c  }{ }&
  $s^0$ & $s^1$ & $s^2$ & $s^3$ & $s^4$ \\
\hline
 $ q_v $ &
 $ A_0 $  &
 $ 4.27 $ & $ 3.096 $ & $ 1.621 $ & $ 0 $ & $ 0 $ \\
\cline{2-7}
 &
 $ A_1 $  &
 $ -4.74 $ & $ -6.90 $ & $ -2.439 $ & $ 0 $ & $ 0 $ \\
\cline{2-7}
 &
 $ A_2 $  &
 $ 2.837 $ & $ 6.46 $ & $ 4.10 $ & $ 0 $ & $ 0 $ \\
\cline{2-7}
 &
 $ B $  &
 $ 0.678 $ & $ -0.0394 $ & $ 0.01758 $ & $ 0 $ & $ 0 $ \\
\cline{2-7}
 &
 $ C $  &
 $ 0.1728 $ & $ -0.02493 $ & $ 0.1451 $ & $ 0 $ & $ 0 $ \\
\hline
 $ g $ &
 $ A $  &
 $ 2.378 $ & $ -4.38 $ & $ 0.585 $ & $ 8.34 $ & $ -9.92 $ \\
\cline{2-7}
 &
 $ B $  &
 $ -0.479 $ & $ -0.607 $ & $ 1.458 $ & $ -6.03 $ & $ 9.33 $ \\
\cline{2-7}
 &
 $ C $  &
 $ 17.06 $ & $ 4.96 $ & $ 24.97 $ & $ -158.2 $ & $ 295.4 $ \\
\cline{2-7}
 &
 $ A'_{0} $  &
 $ 0.2992 $ & $ 1.179 $ & $ -1.915 $ & $ 7.26 $ & $ -18.39 $ \\
\cline{2-7}
 &
 $ A'_{1} $  &
 $ -0.1600 $ & $ -1.114 $ & $ 2.939 $ & $ -6.66 $ & $ 19.23 $ \\
\cline{2-7}
 &
 $ B' $  &
 $ -0.483 $ & $ 0.755 $ & $ -3.80 $ & $ 10.75 $ & $ -19.93 $ \\
\cline{2-7}
 &
 $ C' $  &
 $ 1.297 $ & $ -0.1669 $ & $ 1.906 $ & $ -2.057 $ & $ 0 $ \\
\hline
 $ q_{sea} $ &
 $ A $  &
 $ 3.34 $ & $ -5.61 $ & $ 50.0 $ & $ -220.7 $ & $ 302.8 $ \\
\cline{2-7}
 &
 $ B_0 $  &
 $ -0.2402 $ & $ -0.409 $ & $ 2.263 $ & $ -10.50 $ & $ 14.87 $ \\
\cline{2-7}
 &
 $ B_1 $  &
 $ 8.79 $ & $ -8.86 $ & $ 164.0 $ & $ -712. $ & $ 973. $ \\
\cline{2-7}
 &
 $ C $  &
 $ 9.16 $ & $ 9.29 $ & $ -278.4 $ & $ 1175. $ & $ -1592. $ \\
\hline
 $ \delta c_v $ &
 $ A_0 $  &
 $ 0 $ & $ 0.1219 $ & $ 6.20 $ & $ -25.04 $ & $ 30.98 $ \\
\cline{2-7}
 &
 $ A_1 $  &
 $ 0 $ & $ 1.913 $ & $ -76.9 $ & $ 318. $ & $ -392. $ \\
\cline{2-7}
 &
 $ A_2 $  &
 $ 0 $ & $ -7.16 $ & $ 250.3 $ & $ -1062. $ & $ 1308. $ \\
\cline{2-7}
 &
 $ A_3 $  &
 $ 0 $ & $ 3.19 $ & $ -230.1 $ & $ 1012. $ & $ -1250. $ \\
\cline{2-7}
 &
 $ B $  &
 $ 0.499 $ & $ 3.47 $ & $ -15.26 $ & $ 19.67 $ & $ 0 $ \\
\cline{2-7}
 &
 $ C $  &
 $ 0.329 $ & $ 8.24 $ & $ -38.0 $ & $ 46.3 $ & $ 0 $ \\
\hline
 $ \delta c_{sea} $ &
 $ A $  &
 $ 0 $ & $ -0.0499 $ & $ 0.1026 $ & $ -0.0787 $ & $ 0 $ \\
\cline{2-7}
 &
 $ B_0 $  &
 $ -0.361 $ & $ -0.576 $ & $ 0.2257 $ & $ 0 $ & $ 0 $ \\
\cline{2-7}
 &
 $ B_1 $  &
 $ 7.68 $ & $ -8.83 $ & $ 3.88 $ & $ 0 $ & $ 0 $ \\
\cline{2-7}
 &
 $ C $  &
 $ 2.548 $ & $ 0.691 $ & $ -8.70 $ & $ 10.65 $ & $ 0 $ \\
\end{tabular}
\end{table}

\begin{figure}
\caption { The data and the theoretical predictions for the photon
structure
  function $F_2^{\gamma}/\alpha$.
  The vertical axes show $F_2^{\gamma}/\alpha$ and the horizontal
axes represent
  the scaling variable $x$.
  The data points with a cross mark are not used for the fit,
  according to the selection criterion Eq.~(\protect\ref{3.1}).
  The OPAL data points are
  obtained by removing the direct charm quark contribution. We hence
  drop from the theoretical curves the
  $\gamma^*\gamma\rightarrow c\bar c$ contributions. The remaining
  structure function data contain all hadronic final states. \quad
  (a) The best fits with WHIT1--WHIT3 gluon distributions.
  (b) The best fits with WHIT4--WHIT6 gluon distributions.  }
\label{F3.1}
\end{figure}

\begin{figure}
\caption{The deviation of each data point from the best fit,
  $F_2^\gamma(x,Q^2)_{\rm fit} - F_2^\gamma(x,Q^2)_{\rm data}$
  divided by the error
  $\sigma(F_2^\gamma(x,Q^2)_{\rm data})$.
  A data point with smaller $x$ value is placed lower in each data
  set with a common $\langle Q^2 \rangle$.
  The data points with a simple cross mark are removed from the fit
  by the criterion Eq.~(\protect\ref{3.1}).  }
\label{F3.2}
\end{figure}

\begin{figure}
\caption{ The $C_g$ dependence of the total $\chi^2$ for fixed
$A_g$'s when
  the standard valence-quark
  distributions are taken
  for $A_g$ = 0.5 (solid line) and 1.0 (dashed line).
  The large square, diamond and cross marks are the minimal $\chi^2$
  values as obtained by tuning  the valence-quark distributions.}
\label{F3.3}
\end{figure}

\begin{figure}
\caption{  Gluon distributions at $Q^2$ = 4, 20 and 100 GeV$^2$.
  The top 3 figures are for WHIT1 (solid), WHIT2 (dotted) and WHIT3
  (dashed), the middle 3 figures are WHIT4 (solid), WHIT5 (dotted)
and
  WHIT6 (dashed), and the bottom 3 figures are
GRV\protect\cite{GRV}(solid),
  DG\protect\cite{DG} (dotted) and LAC1\protect\cite{ACL}(dashed) for
comparison.  }
\label{F4.1}
\end{figure}

\begin{figure}
\caption{%
  Predictions for the charm quark distribution in the photon,
calculated by
  the QPM (solid lines) and by the massive inhomogeneous
  AP equations (dash-dotted lines).
  The valence-up-quark distributions
  of WHIT1-WHIT3 (dashed lines) and WHIT4-WHIT6 (dotted lines) are
  also shown  for comparison.   }
\label{Fig:charm}
\end{figure}

\begin{figure}
\caption{  The differential cross section of the process
  $e^+e^-\rightarrow e^+e^-c\bar c$
  at TRISTAN energy ($\protect\sqrt{s}=58 {\rm GeV}$),
  evaluated exactly and via EPA: (a) near the charm quark pair
threshold
  ($\protect\sqrt{\hat s}=m_{cc}=4{\rm GeV}$) and; (b) at far above
  the threshold ($\protect\sqrt{\hat s}=m_{c\bar c}=15{\rm GeV}$).
  Vertical bars indicate errors of numerical integration of the exact
  matrix elements.
  We set $m_c$=1.5 GeV and $\alpha$=1/137.}
\label{Fig:EPACa}
\end{figure}
\begin{figure}
\caption{ The invariant mass distribution of charm quark pair
  in the process $e^+e^-\rightarrow e^+e^-c\bar c$ evaluated exactly
and via
  EPA at $\protect\sqrt{s}=58{\rm GeV}$.
  We set $m_c=1.5$ GeV, and $\alpha=1/137$}.
\label{Fig:EPACc}
\end{figure}

\begin{figure}
\caption{The leading order prediction for the total cross section of
the
  inclusive process $e^+e^-\rightarrow e^+e^-c\bar cX$.
  The contributions from the resolved photon processes depend on the
parton
  distributions WHIT1 to 6.
  The curves are obtained by setting $m_c=1.5$GeV,
  $\alpha=1/137$ and $\Lambda_4=0.4$ GeV for the strong coupling,
  and by requiring
  the invariant mass $W$ of the hadron system to satisfy
  $W\geq 2m_D=3.74$ GeV.
  The vertical bars attached to the WHIT1, WHIT4 and WHIT6
predictions
  indicate the
  dependence of the cross sections on the charm mass $m_c$ between
  1.3 GeV and 1.7 GeV.}
\label{Fig:EPA}
\end{figure}
\begin{figure}
\caption{ The leading order predictions for
  the total cross sections of the inclusive process
  $e^+e^-\rightarrow e^+e^-c\bar cX$,
  where  contributions of the direct and resolved photon processes
are shown
  separately.
  The parton distributions of
  WHIT1 (a) and WHIT6 (b) are used to calculate the resolved
  photon contributions. We set $m_c=1.5$GeV, $\alpha=1/137$ and
  $\Lambda_4=0.4$ GeV.
  The vertical bars indicate the dependence of the cross sections
  on the charm mass $m_c$ between 1.3 GeV and 1.7 GeV.}
\label{Fig:EPAa}
\end{figure}
\ifFigureInclude{
\newpage
%

\begin{center}
  {Fig.~\ref{F3.1}~~(a)}\\
\vspace*{10cm}
\epsfxsize=14.5cm
\leavevmode
\epsfbox{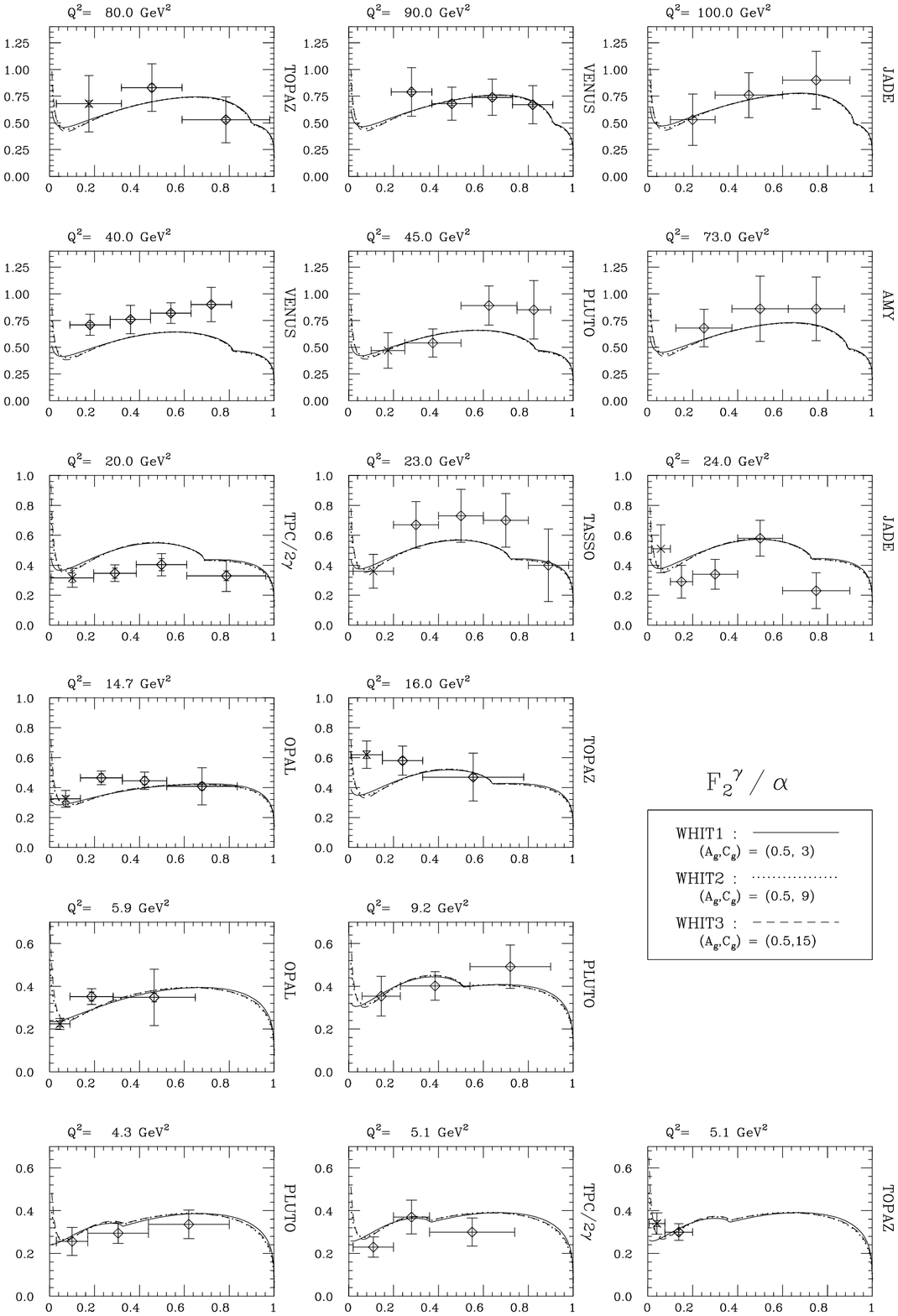}
\end{center}

\newpage

\begin{center}
  {Fig.~\ref{F3.1}~~(b)}\\
\vspace*{10cm}
\leavevmode
\epsfxsize=14.5cm
\epsfbox{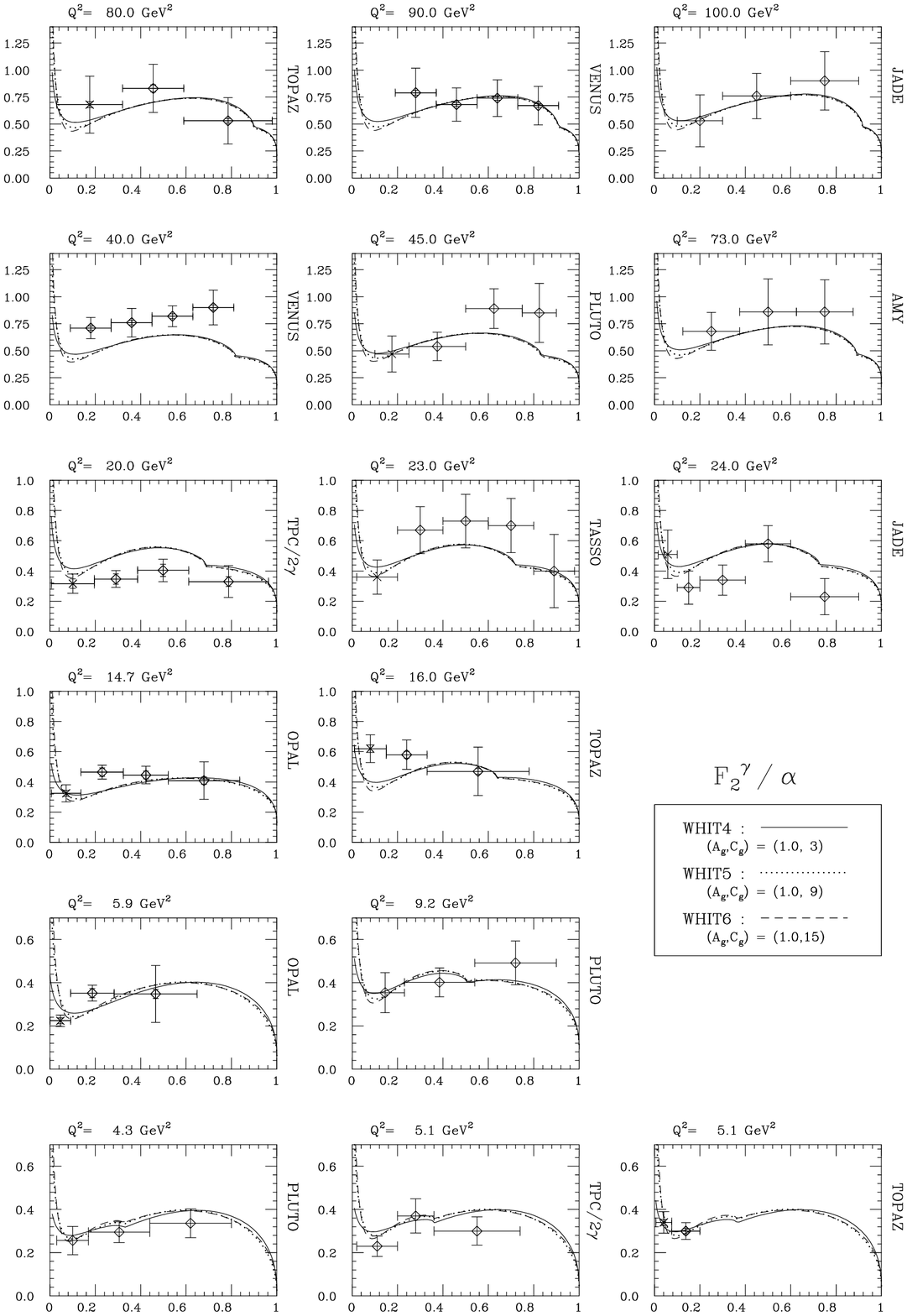}
\end{center}

\newpage

\begin{center}
  {Fig.~\ref{F3.2}}\\
\vspace*{10cm}
\leavevmode
\epsfxsize=15.5cm
\epsfbox{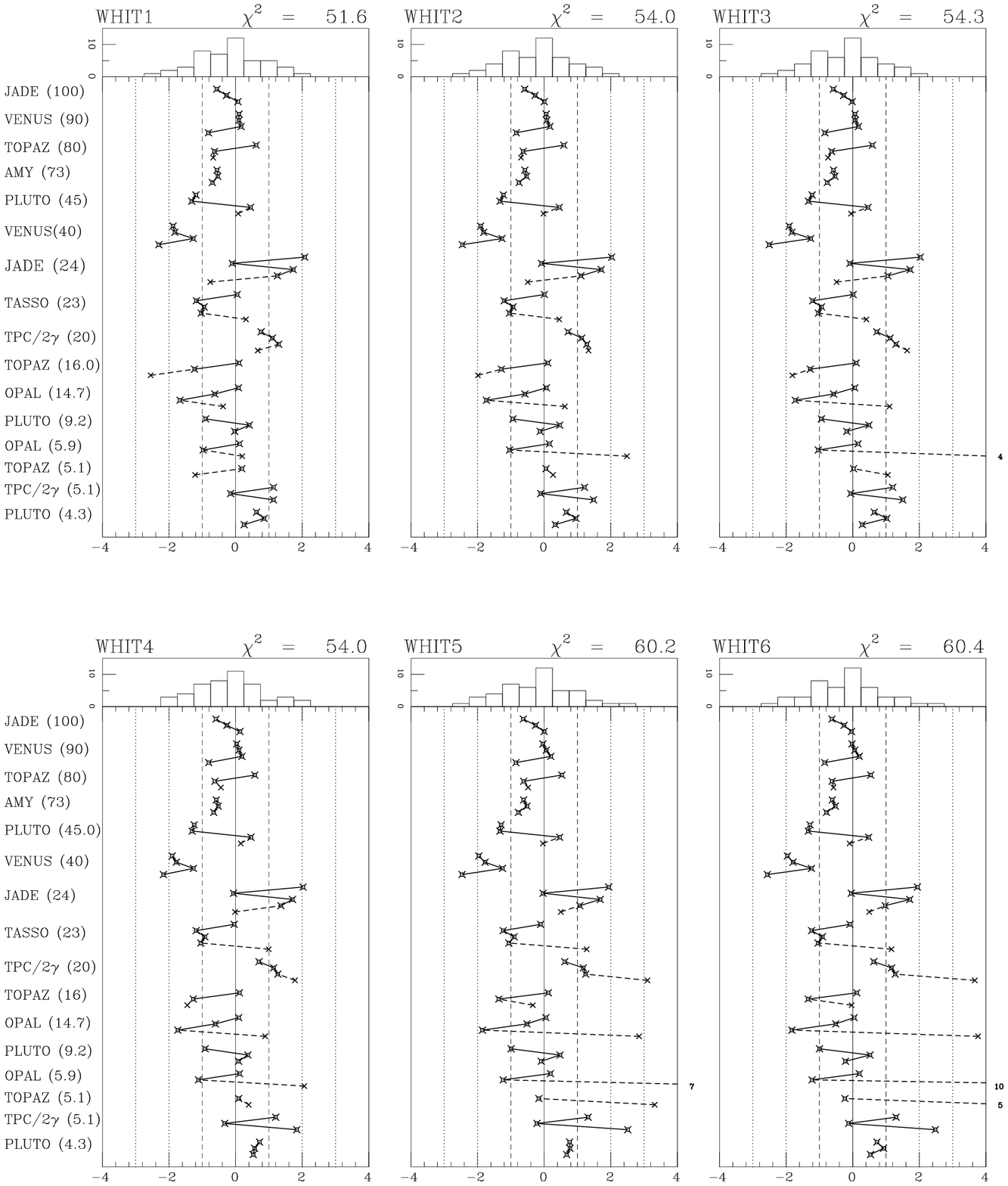}
\end{center}

\newpage

\begin{center}
  {Fig.~\ref{F3.3}}\\
\vspace*{10cm}
\leavevmode
\epsfxsize=14.5cm
\epsfbox{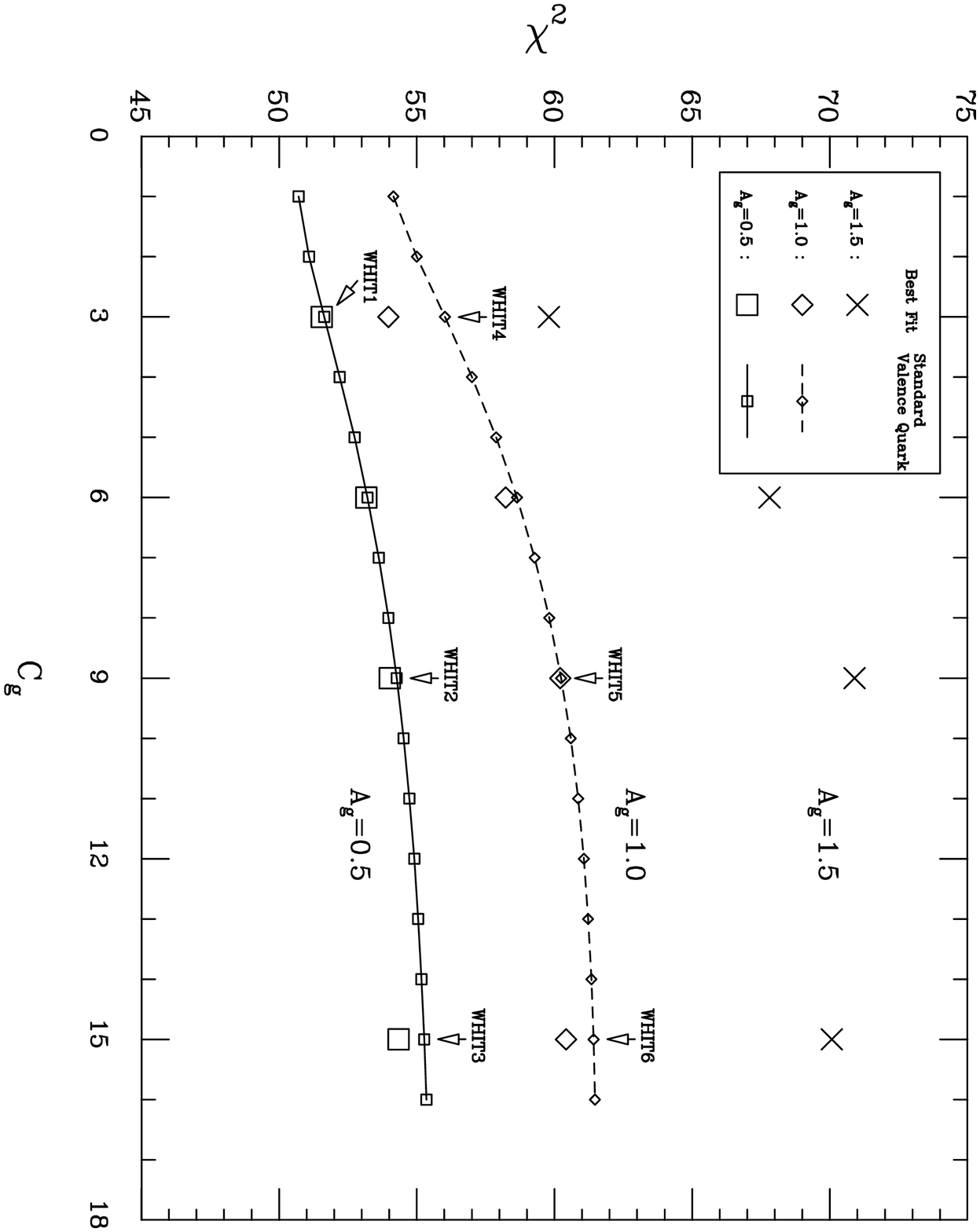}
\end{center}

\newpage

\begin{center}
  {Fig.~\ref{F4.1}}\\
\vspace*{10cm}
\leavevmode
\epsfxsize=14.5cm
\epsfbox{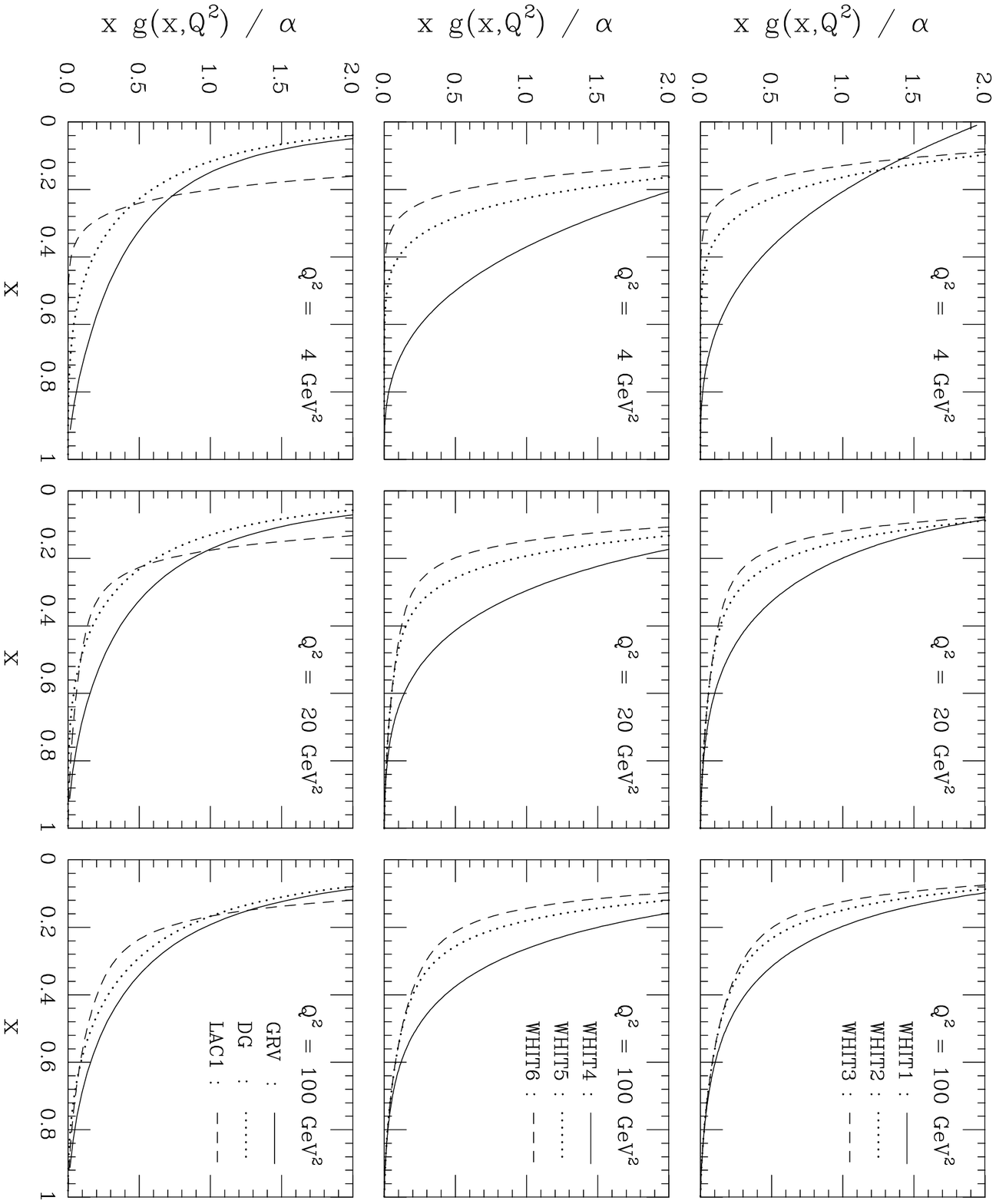}
\end{center}

\newpage

\begin{center}
  {Fig.~\ref{Fig:charm}}\\
\vspace*{10cm}
\leavevmode
\epsfxsize=14.5cm
\epsfbox{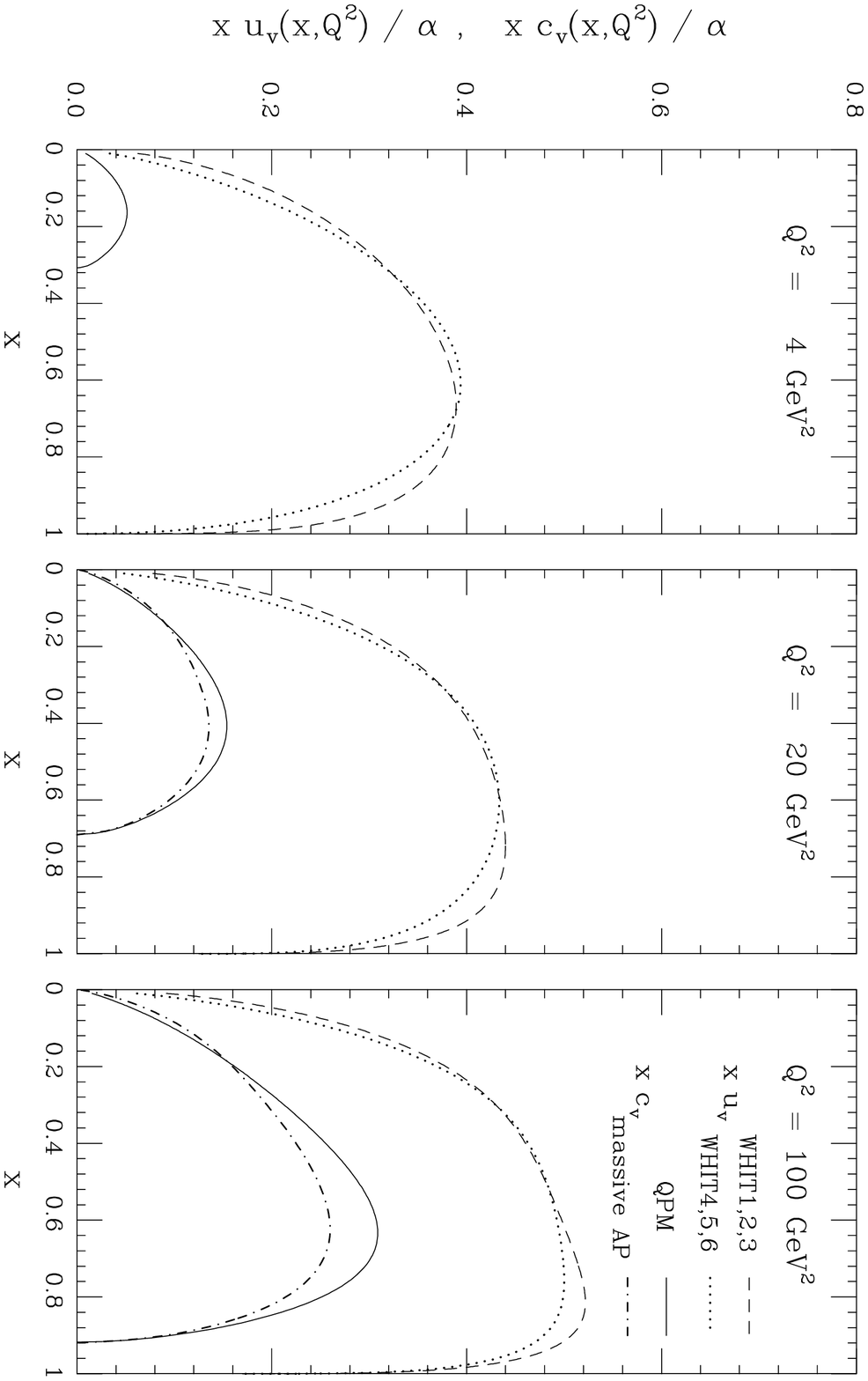}
\end{center}

\newpage

\begin{center}
  {Fig.~\ref{Fig:EPACa}~~(a)}\\
\vspace*{10cm}
\leavevmode
\epsfxsize=14.5cm
\epsfbox{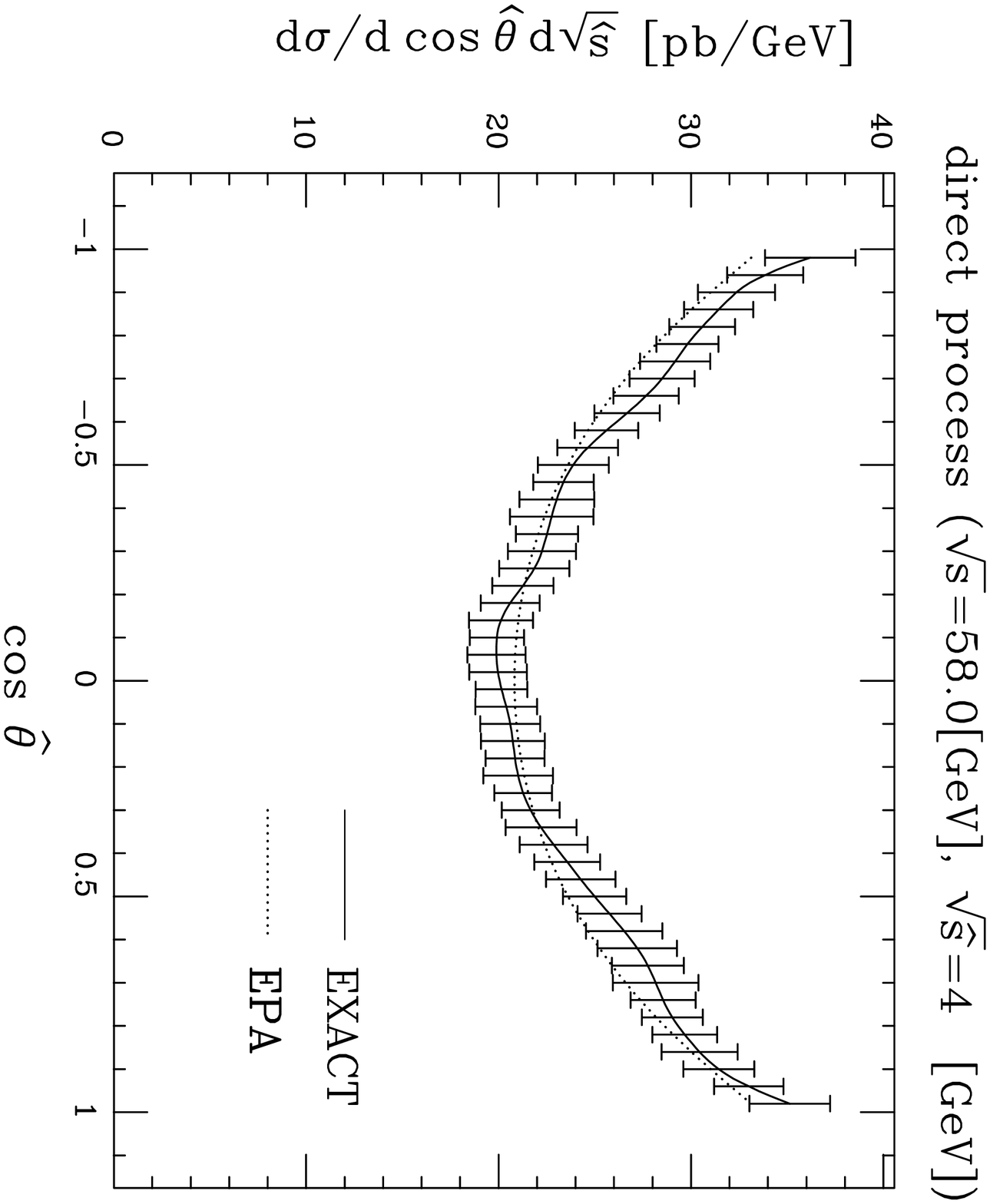}
\end{center}

\newpage

\begin{center}
  {Fig.~\ref{Fig:EPACa}~~(b)}\\
\vspace*{10cm}
\leavevmode

\epsfxsize=14.5cm
\epsfxsize=14.5cm
\epsfbox{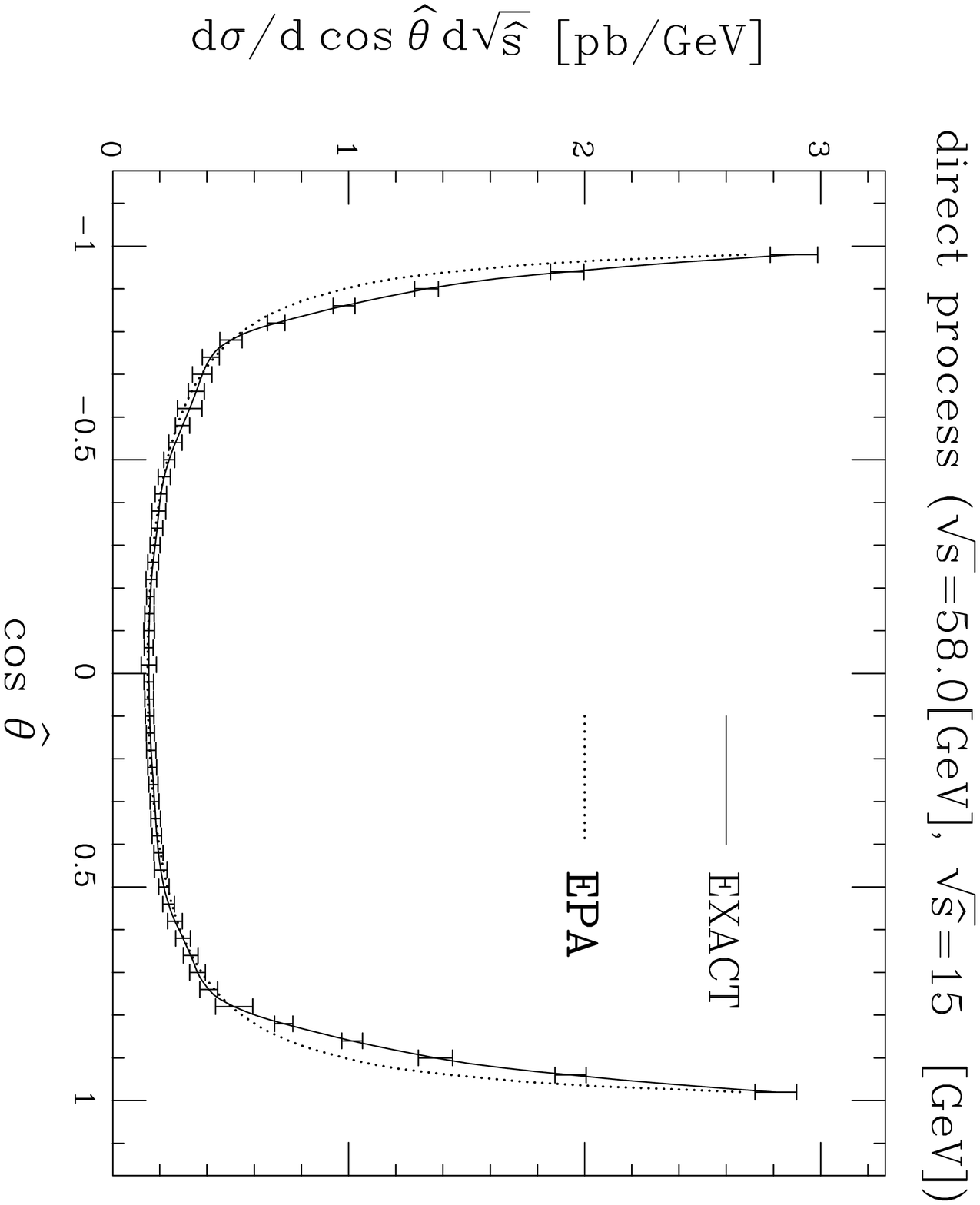}
\end{center}

\newpage

\begin{center}
  {Fig.~\ref{Fig:EPACc}}\\
\vspace*{10cm}
\leavevmode
\epsfxsize=14.5cm
\epsfbox{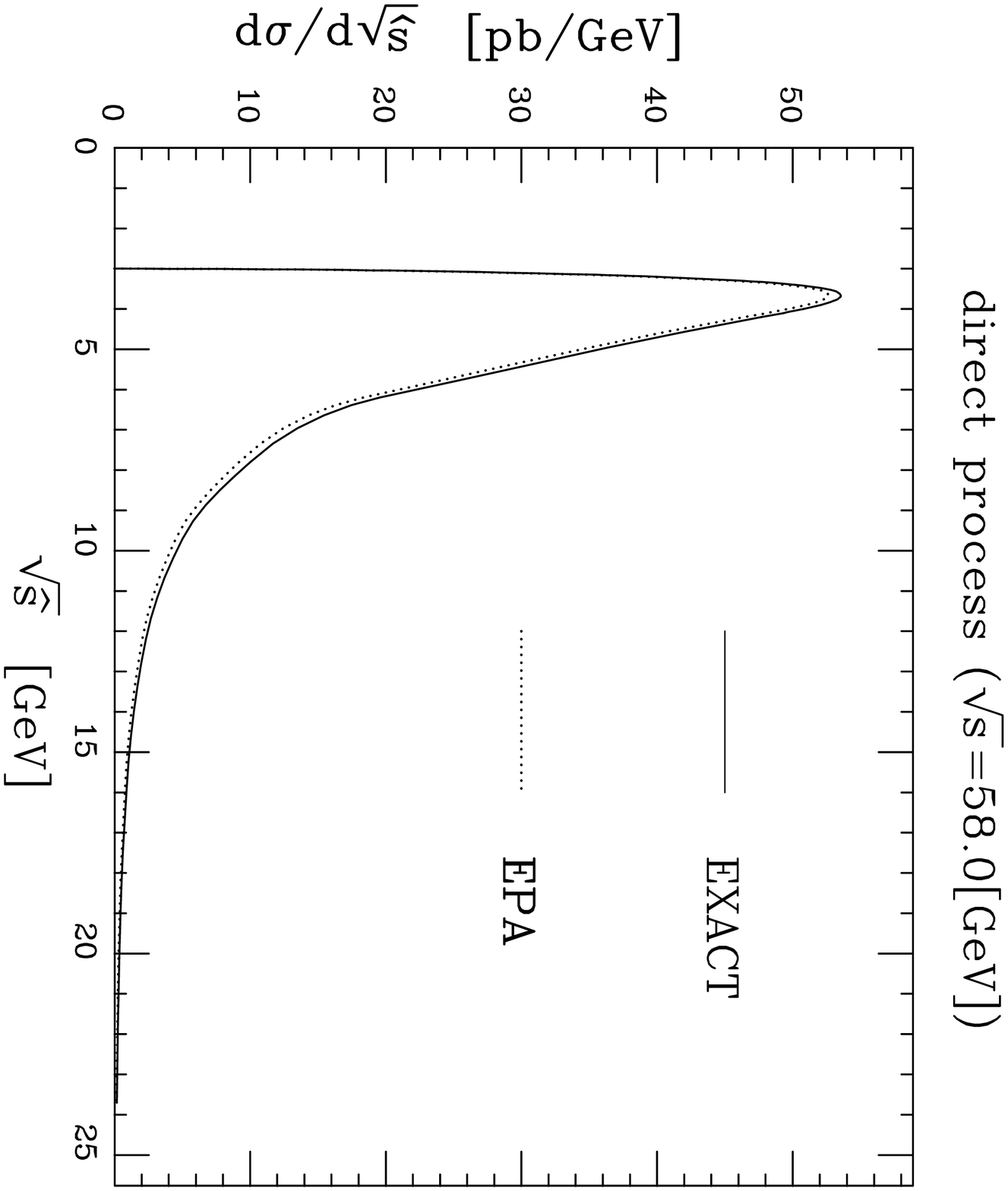}
\end{center}

\newpage

\begin{center}
  {Fig.~\ref{Fig:EPA}}\\
\vspace*{10cm}
\leavevmode
\epsfxsize=14.5cm
\epsfbox{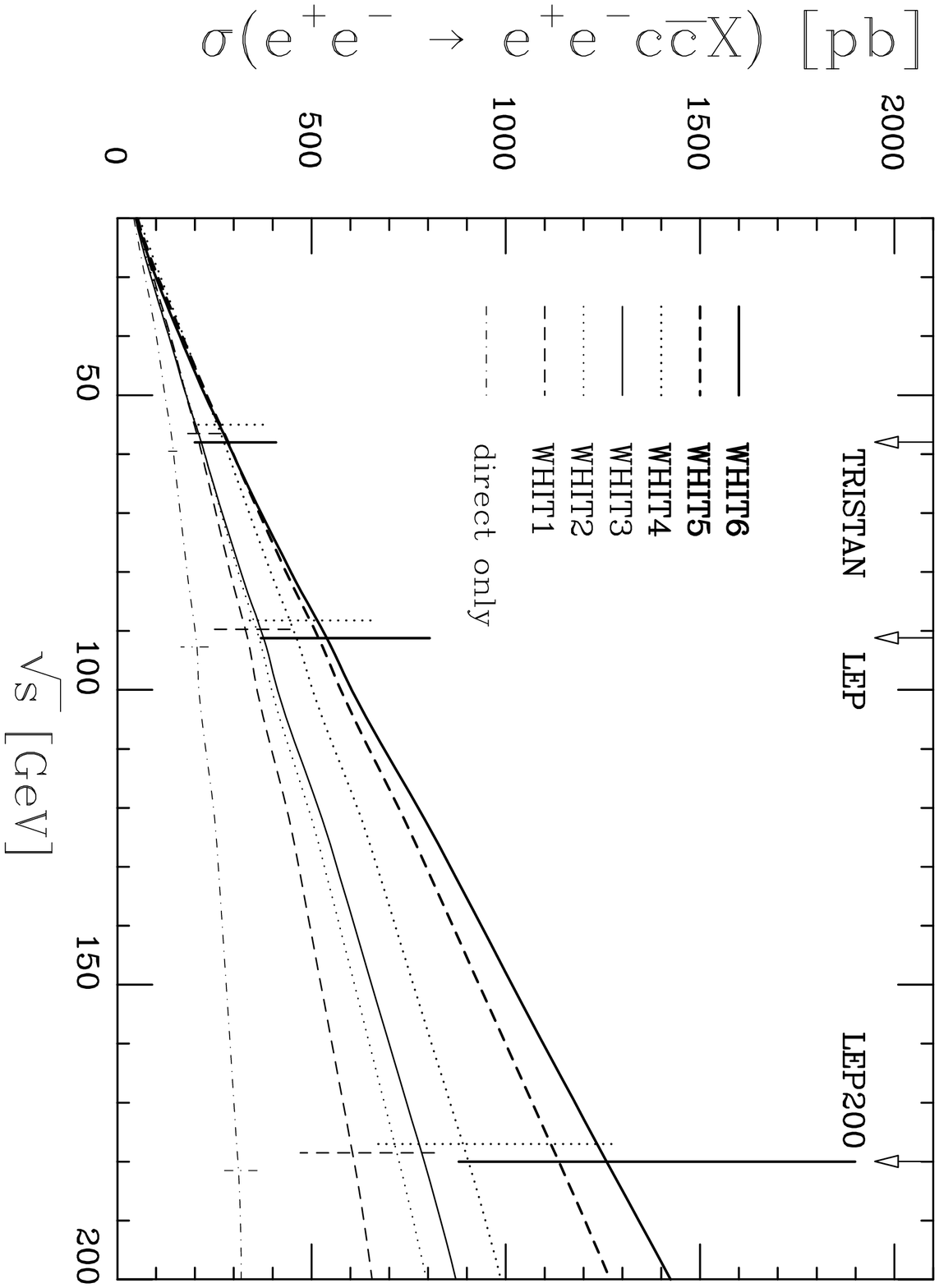}
\end{center}

\newpage

\begin{center}
  {Fig.~\ref{Fig:EPAa}~~~(a)}\\
\vspace*{10cm}
\leavevmode
\epsfxsize=14.5cm
\epsfbox{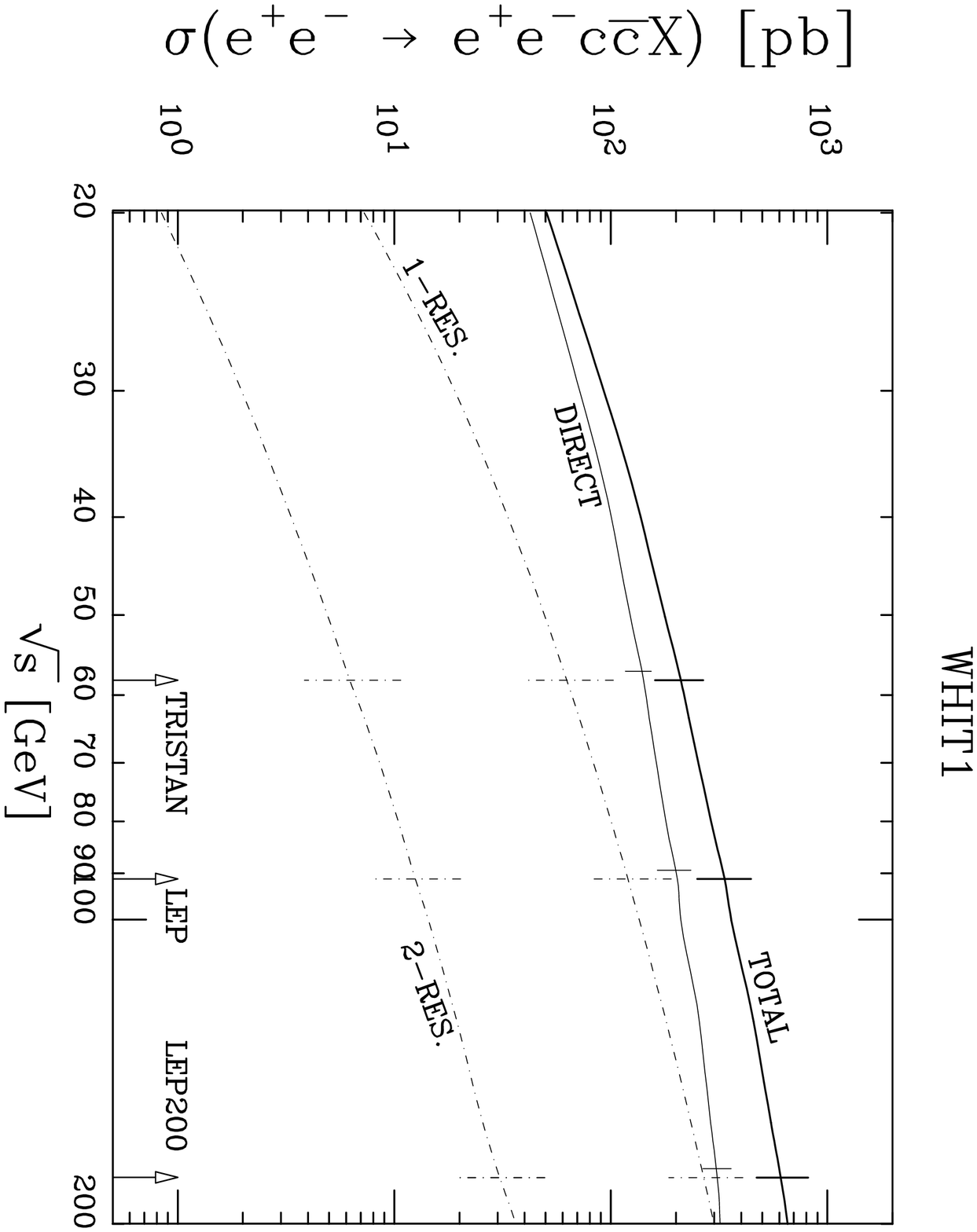}
\end{center}

\newpage

\begin{center}
  {Fig.~\ref{Fig:EPAa}~~~(b)}\\
\vspace*{10cm}
\leavevmode
\epsfxsize=14.5cm
\epsfbox{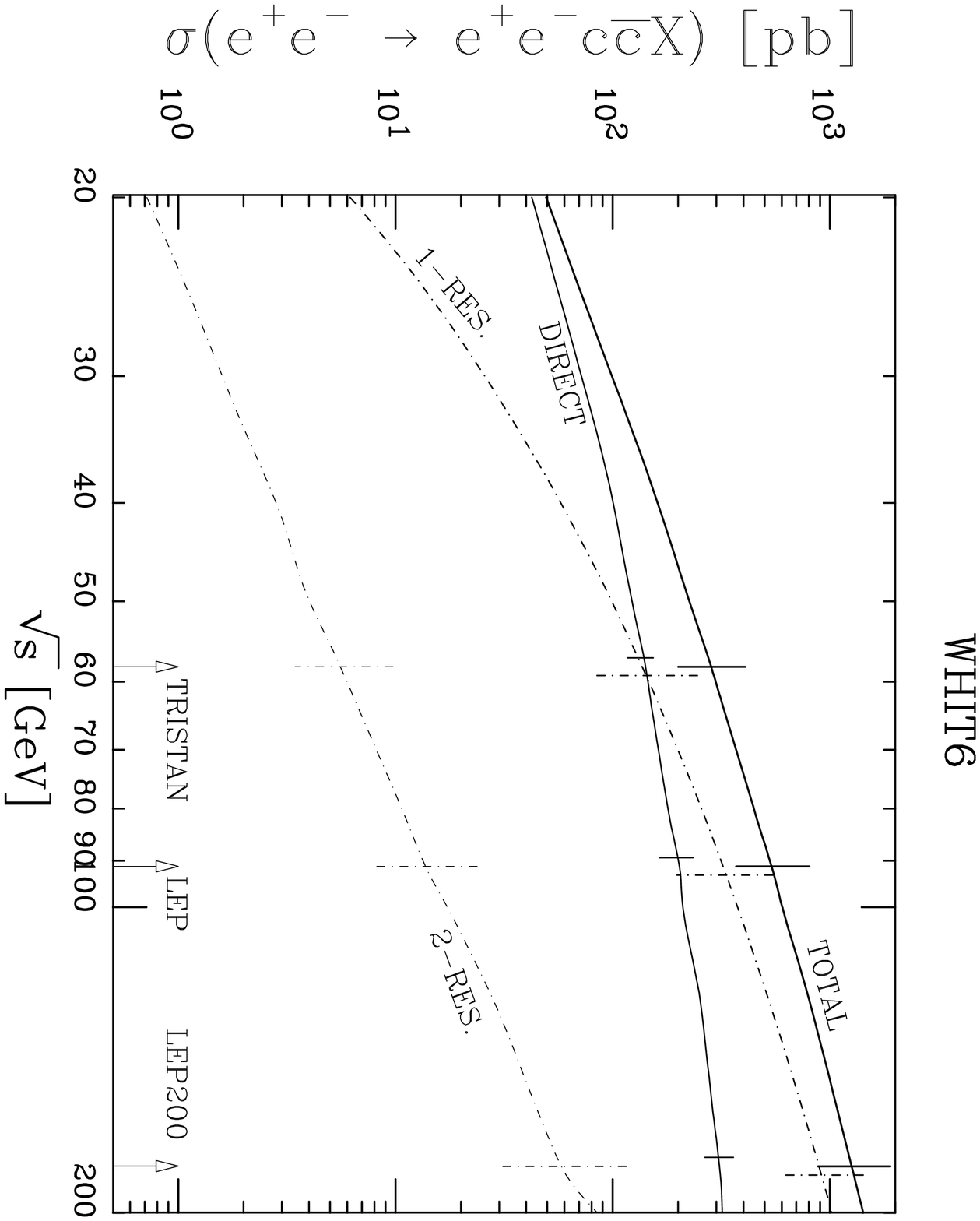}
\end{center}
}
\fi
\end{document}